\documentclass[aps,prresearch,reprint,superscriptaddress,longbibliography]{revtex4-2}
\usepackage{xcolor}
\usepackage{multirow}
\usepackage{array}
\usepackage{makecell} 
\usepackage{braket}
\usepackage{graphicx}
\usepackage{amsmath}
\usepackage{amssymb}
\usepackage{hyperref}
\usepackage{enumitem}
\usepackage{float}
\setcellgapes{3pt} 
\makegapedcells     

\begin{document}

\title{Entanglement without Quantum Mechanics: Operational Constraints on the Quantum Signature}
\author{Samuel Schlegel}
\email{samuel.schlegel@univie.ac.at}
\affiliation{University of Vienna, Faculty of Physics, Vienna Center for
Quantum Science and Technology, Boltzmanngasse 5,
Vienna 1090, Austria}
\author{Borivoje Daki\'c}
\affiliation{University of Vienna, Faculty of Physics, Vienna Center for
Quantum Science and Technology, Boltzmanngasse 5,
Vienna 1090, Austria}
\affiliation{Institute for Quantum Optics and Quantum Information (IQOQI),
Austrian Academy of Sciences, Boltzmanngasse 3, Vienna 1090,
Austria}
\author{Flavio Del Santo} 
\affiliation{University of Vienna, Faculty of Physics, Vienna Center for
Quantum Science and Technology, Boltzmanngasse 5,
Vienna 1090, Austria}
\affiliation{University of Geneva, Group of Applied Physics, Rue de l'\'Ecole-de-M\'edecine 21, Geneva 1211, Switzerland}
\affiliation{Constructor University, Bremen, Germany}

\begin{abstract}
Quantum entanglement is defined as nonseparability of density operators, yet experimental phase-space statistics alone cannot distinguish classical from quantum sources, so classical data can be used to violate entanglement inequalities. Imposing uncertainty bounds, or full tomography, can map the same data onto positive, entangled density operators, yielding a hybrid regime in which classical and quantum models coincide. Thus, certifying genuine quantum entanglement requires additional tests, such as Wigner negativity, incompatibility beyond phase-space measurements, or nonlinear dynamics.
\end{abstract}

\maketitle

\section{Introduction}
According to the textbook narrative, the emergence of quantum mechanics shattered our classical, intuitive picture of reality. However, much of this narrative blends together the framework in which a theory is formulated and the physical content of the underlying theory. Entanglement, in particular, is often regarded as the hallmark of quantum mechanics, as it exhibits correlations with no classical explanation: already Schr\"odinger, who introduced the concept, called entanglement ``not \textit{one} but rather \textit{the} characteristic trait of quantum mechanics'' \cite{schrodinger1935discussion}. 

In recent years, novel attempts have been put forward to scale down the fundamental difference between classical and quantum theory by showing that certain features regarded as genuinely quantum can already be found in classical models if one imposes epistemic constraints \cite{spekkens_defense_2007, catani1, catani2023interference, catani2023aspects, Fankhauser_2025}, fundamental indeterminacy due to finite information \cite{del2019physics, santo_which_2025}, classical--anti-classical toy models \cite{chiribella_bell_2024}, or using operational probabilistic theories \cite{dariano_classical_2020}. In particular, Refs. \cite{chiribella_bell_2024, santo_which_2025, dariano_classical_2020} have proposed ways to construct analogues of entanglement in their respective proposed classical models. 
Parallel to these developments, the debate on “classical entanglement” in optics has highlighted the formal equivalence between quantum entanglement and the nonseparable coupling of different degrees of freedom---e.g., polarization and spatial mode---within a single classical electromagnetic field \cite{spreeuw_classical_1998, aiello_quantum_2015, forbes_chapter_2019}. However, these correlations are simply  mathematical analogies and are operationally distinct from entanglement, which becomes relevant when it involves distant subsystems \cite{karimi_classical_2015, korolkova_operational_2024, peres1997quantum}.
 
Yet, while these approaches can be insightful to understand certain theoretical features, they typically rely on extending or modifying the underlying theoretical framework, remaining toy theories, proofs of principles to point out that entanglement \textit{can} be constructed in classical frameworks. In contrast, our analysis is carried out strictly within \emph{the bounds of classical and quantum mechanics}. We show that certain features commonly associated with one theory can be reproduced within the other when the \emph{same physics is expressed in a different representation}. Just as quantum states can exhibit classical phase-space features (most notably a Wigner function that takes non-negative values), classical states, when translated into the Hilbert-space language, can display apparent entanglement-like correlations and even satisfy uncertainty relations and positivity (of density operators).

Our main goal is to study how far classical physics—when translated into the quantum formalism—can give rise to signatures typically regarded as genuinely quantum, such as entanglement. In quantum theory, entanglement is defined as nonseparability of density operators on a tensor-product Hilbert space. Yet in practice it is inferred only from experimental data, which do not \emph{a priori} reveal whether the underlying system is classical or quantum. This motivates an operational viewpoint in which the same measurement record may be analysed in different mathematical languages, independently of the system’s ontic status. In particular, the Wigner–Weyl correspondence~\cite{gerry_introductory_2023} translates between phase-space descriptions and Hilbert-space operators, allowing one to represent phase-space statistics as operators (and conversely). Within this comparative setting, we show that classical states can display entanglement-like correlations certified by covariance-based witnesses~\cite{duan_inseparability_2000}, and may do so while still satisfying covariance uncertainty constraints~\cite{serafini_quantum_2023, weedbrook_gaussian_2012}; we refer to this as \emph{representational entanglement}. Moreover, applying standard quantum-state reconstruction techniques to purely classical quadrature data can yield positive (density-operator) reconstructions that are nevertheless entangled, giving rise to a \emph{hybrid regime} in which genuinely quantum states and classical phase-space models reproduce the same phase-space statistics and are therefore indistinguishable by such measurements alone. Only when the observer’s operational capabilities are extended—enabling incompatibility tests or other nonclassical signatures such as Wigner negativity—can one certify entanglement that admits no classical description.

Let us state precisely the conceptual advance with respect to existing literature. Prior approaches construct entanglement analogues within modified or restricted theoretical frameworks~\cite{spekkens_defense_2007, dariano_classical_2020, chiribella_bell_2024}, characterise the relation between Wigner positivity and Gaussianity for single-mode states~\cite{Mandilara_2009, genoni_detecting_2013}, or reconstruct the Gaussian sector of quantum mechanics from epistemically restricted Liouville mechanics~\cite{bartlett_reconstruction_2012}. Our contribution differs in three regards. First, we exhibit the false-positive direction, as covariance-based separability criteria in routine experimental use can certify ``entanglement'' for operators that fail positivity and hence correspond to no quantum state at all. Whereas second-moment witnesses are known to miss genuine entanglement for non-Gaussian states~\cite{gomes_quantum_2009, hertz_detection_2016}, the failure mode identified here is the converse one of false certification. Second, we extend the characterization of the overlap to the bipartite non-Gaussian regime, providing explicit examples of entangled, Wigner-positive states beyond the Gaussian sector. Third, we organise these observations into an operational hierarchy of nonseparability tied to concrete measurement capabilities, from covariance data through tomographic positivity checks to direct probes of Wigner negativity.

\begin{figure}
    \centering
    \includegraphics[width=1\linewidth]{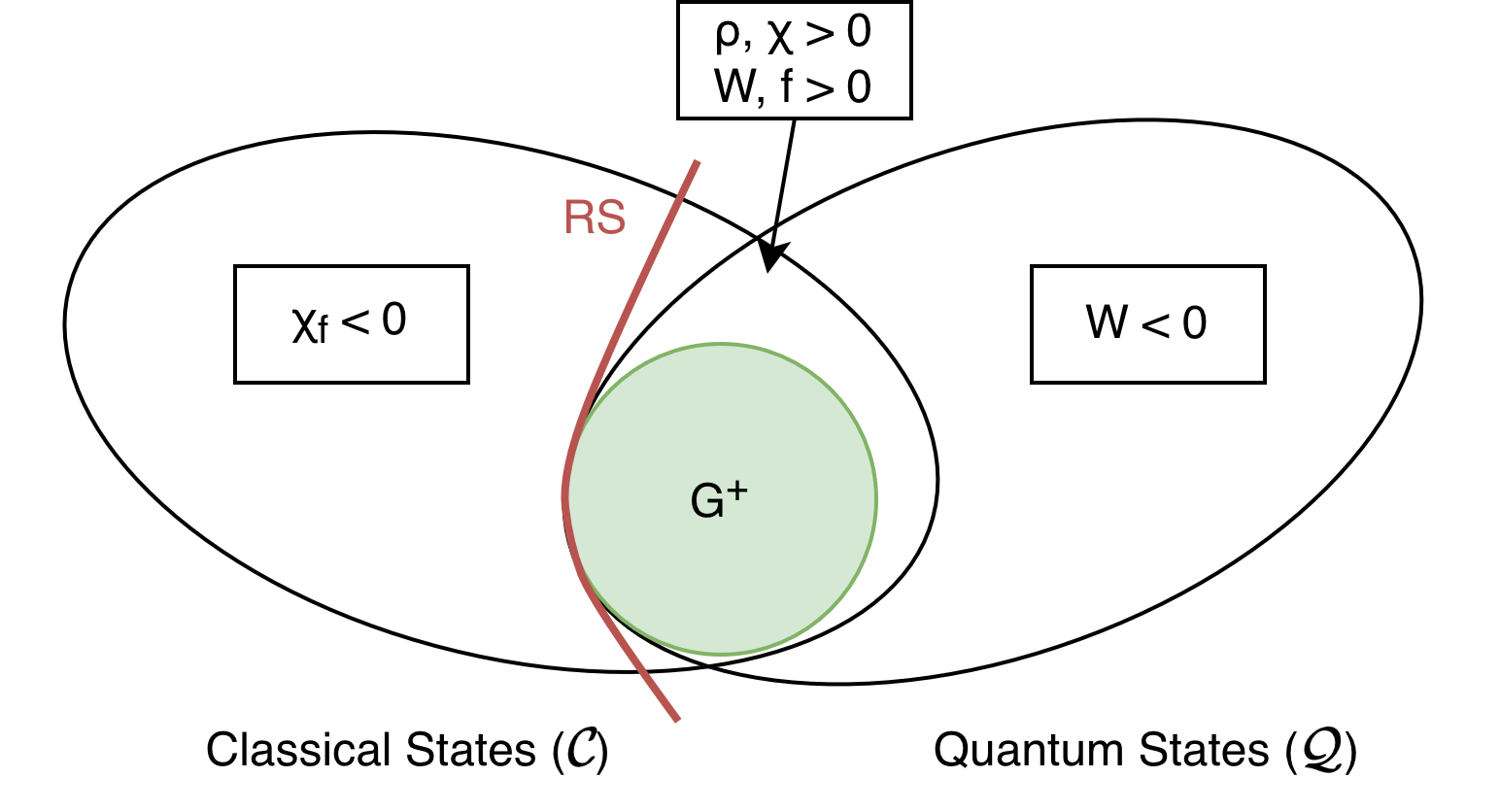}
    \caption{Overlap of classical and quantum state spaces in the Wigner--Weyl representation. States in the intersection are operationally indistinguishable when access is restricted to phase-space (quadrature) statistics.}
    \label{fig:statespace}
\end{figure}

\section{Classical States in Hilbert Space}
As is well known, classical physics is customarily represented in a real phase space, whereas quantum theory is formulated in a complex Hilbert space. Classical states are described by probability distributions on phase space (with pure states corresponding to Dirac delta functions), while quantum states are represented by density operators (with pure states corresponding to rays in Hilbert space). However, this choice of formalism is somewhat arbitrary: it is, in fact, possible to express both classical and quantum mechanics within a common framework using the Wigner--Weyl formalism~\cite{weyl_quantenmechanik_1927,moyal_quantum_1949,case_wigner_2008}. 

On the one hand, quantum states $\hat{\rho}$ can be represented as quasiprobability distributions in phase space via the Wigner transform
\begin{equation} 
\label{eq:wigner_trans} 
W_{\rho}(q,p) = \frac{1}{2 \pi \hbar} \int_{-\infty}^\infty ds \ e^{-\frac{i}{\hbar}ps} \bra{q+\frac{s}{2}}\,\hat{\rho}\, \ket{q- \frac{s}{2}}, 
\end{equation}
which reduces expectation values to phase-space integrals and maps simple projectors onto delta distributions~\cite{hillery_distribution_1984}. The resulting Wigner function $W_\rho$ resembles a probability density but can attain negative values; such negativity is usually taken as a signature of nonclassicality~\cite{kenfack_negativity_2004}.

On the other hand, our main goal is to represent classical states in Hilbert space via the inverse map---the Weyl transform. For a normalized classical distribution $f(q,p)$ we define the corresponding operator
\begin{equation}
\label{eq:weyl_trans}
    \hat{\chi}_f 
    = W^{-1}[f]
    = \int_{-\infty}^{\infty} dq \int_{-\infty}^{\infty} dp\;
      f(q,p)\,\hat{\Delta}(q,p),
\end{equation}
where $\Delta$ denotes the Stratonovich--Weyl kernel \footnote{$\hat{\Delta}(q,p) = \protect\int_{-\infty}^{\infty} ds\; e^{\frac{i}{\hbar}ps}\, \ket{q+\tfrac{s}{2}}\bra{q-\tfrac{s}{2}}$} \cite{stratonovich1957distributions}.

Since our primary interest is to study the nature of correlations, we shall move directly to the bipartite mapping. A classical phase-space density  $f(q_1,p_1,q_2,p_2)$ is mapped to
\begin{equation}
    \hat{\chi}_f
    = \int dq_1\,dp_1\,dq_2\,dp_2\;
      f(q_1,p_1,q_2,p_2)\,
      \hat{\Delta}(q_1,p_1)\otimes\hat{\Delta}(q_2,p_2).
\end{equation}
This construction yields a one-to-one correspondence between normalized classical phase-space densities $f$ and trace-one operators $\chi_f$ on Hilbert space, although these operators are not guaranteed to be positive, and thus do not in general represent valid quantum states. We denote the set of such ``classical'' operators by $\mathcal{C}$, and the set of genuine quantum states (density operators) by $\mathcal{Q}$.  An operator $\hat{\chi}$ belongs to $\mathcal{C}$ if and only if its Wigner representation is everywhere non-negative (equivalently, it arises from a classical probability density $f$ via the Weyl transform), while an operator $\hat{\rho}$ belongs to $\mathcal{Q}$ if and only if it is positive semidefinite.

An important consequence is that neither state space is contained in the other: $\mathcal{C}$ and $\mathcal{Q}$ are distinct, partially overlapping sets (as illustrated schematically in Fig. \ref{fig:statespace}). States in the intersection $\mathcal{C} \cap \mathcal{Q}$ (e.g., Gaussian states and, more generally, Wigner-positive states) are operationally indistinguishable when one restricts to phase-space statistics accessible via quadrature (homodyne) measurements, equivalently to expectation values of Weyl-ordered observables, obtained from measurements on identically prepared ensembles. In this regime, phase-space data alone does not reveal whether a given state has a genuinely quantum origin or arises from a classical distribution.

Moreover, some states in this intersection can exhibit entanglement in the usual Hilbert space sense (explicit examples will be given in the next section). A terminological remark is in order here. Being contained in $\mathcal{C}$ refers to the origin of an operator and not to a physical status that would exclude it from $\mathcal{Q}$. Whenever an operator of classical origin satisfies positivity, it is, by the definitions above, a full-fledged quantum state. The statement is therefore not that classical states become quantum, but that some operators of classical origin are simultaneously bona fide quantum states, namely those in $\mathcal{C} \cap \mathcal{Q}$, i.e. quantum states that admit a description as a classical probability distribution in phase space. With this understanding, operators arising from classical distributions can themselves satisfy \emph{both characteristic features of the quantum formalism}:
\begin{enumerate}
    \item[(A)] \emph{Positivity}: $\hat{\chi} \geq 0$ (and $\mathrm{Tr}\,\hat{\chi}=1)$,
    \item[(B)] \emph{Entanglement}: 
    $\displaystyle \hat{\chi} \neq \sum_i p_i\,\hat{\chi}_A^{(i)} \otimes \hat{\chi}_B^{(i)}$,
    with $p_i \geq 0$, $\sum_i p_i =1$, and each $\hat{\chi}_{A/B}^{(i)}$ satisfying (A).
\end{enumerate}
While this possibility for operators of classical origin is clear from the standpoint of the formalism, one should place (A) and (B) on an operational footing suitable for laboratory testing. To this end, we imagine Alice and Bob, each performing phase-space measurements in their respective laboratories, who wish to test whether their shared system can exhibit entanglement (as in many quantum experiments, e.g.\ in nanomechanical systems~\cite{ockeloen-korppi_stabilized_2018, aspelmeyer_cavity_2014}). 

As our discussion shows, operators of classical origin can also exhibit this feature, provided they satisfy (A) and (B).

In practice, however, verifying (A) and (B) may require operationally demanding techniques (such as full state tomography in the worst-case scenario). What is typically done instead is to replace (A) and (B) by weaker, experimentally friendlier \emph{necessary} conditions, such as covariance-based criteria:
\begin{enumerate}
    \item[(A*)] \label{A*}\emph{Uncertainty relations}, exemplified by the Robertson--Schrödinger (RS) relation~\cite{serafini_quantum_2023, weedbrook_gaussian_2012},
    \item[(B*)] \label{B*}\emph{Positive-partial-transpose (PPT) criterion}, which for continuous-variable bipartite systems reduces to the Duan--Simon criterion formulated in terms of second-order moments (see, e.g. \cite{duan_inseparability_2000, simon_peres-horodecki_2000, adesso_entanglement_2007, serafini_quantum_2023})
\end{enumerate}
In the next section, we examine these tests in detail and show how, within the Wigner--Weyl representation, they can diagnose different layers of nonseparability for both classical and quantum states.

\section{Representational Entanglement}
\label{sec:entanglement_classical}
Let us begin with an observer who has only limited access to the system. We focus solely on experimental constraints, such as access only to the second moments of position and momentum, which are nevertheless sufficient to test conditions (A*) and (B*), which are necessary and sufficient for Gaussian states, but are only necessary beyond Gaussianity. From this perspective, separability is judged solely through covariance data. In such a restricted setting, even entirely classical mixtures can appear non-separable once expressed in the quantum formalism. More precisely, the first condition (A*) is implemented by the covariance-based uncertainty relation, also known as the Robertson--Schrödinger inequality
\begin{equation}
\label{eq:RS}
    \Sigma + \frac{i\hbar}{2} \Omega \succeq 0,
\end{equation}
where $\Sigma$ is the covariance matrix, collecting all variances and covariances of the quadrature observables, and $\Omega$ is the symplectic form.

The second criterion, (B*), corresponds to the Peres-Horodecki PPT criterion \cite{simon_peres-horodecki_2000, duan_inseparability_2000}, which in the covariance formalism becomes 
\begin{equation}
\label{eq:PPT}
    \Sigma^\Gamma + \frac{i\hbar}{2} \Omega \succeq 0,
\end{equation}
where $\Sigma^\Gamma$ denotes the covariance matrix of the state after partial transposition \cite{serafini_quantum_2023}. Operationally, partial transposition corresponds to flipping the sign of a subsystem's momentum in phase space. 
In practice, testing both RS and PPT reduces to the measurement of covariance data and computing the smallest symplectic eigenvalues of $\Sigma$ and $\Sigma^\Gamma$ and checking whether they are at least $\hbar/2$. These criteria are extremely useful within the Gaussian sector. In particular, the RS condition cleanly separates Gaussian states that satisfy the uncertainty relations (the $G^+$ region in Fig. \ref{fig:statespace}) from those that do not and are therefore purely classical. Beyond the Gaussian regime, however, they provide only necessary conditions, which opens the door to more subtle behavior: non-Gaussian classical states can satisfy RS and even violate PPT at the covariance level, while their associated operator fails positivity and thus does not correspond to a physical quantum state. To see how this leads to representational artifacts, recall that the Weyl transform maps a classical phase space distribution $f(z_1, z_2)$, where $z_{1/2} = (q_{1/2}, p_{1/2})$, to a Hilbert space operator $\hat{\chi}_f$. Correlations in $f$ can result in an operator that looks nonseparable in the Hilbert space sense (in a sense of (B*)), even though $\hat{\chi}_f$ is not positive. In other words, classical correlations can mimic the structure of entanglement if we look only at a restricted slice of information (like covariances) and ignore the full operator spectrum.

We stress that the mathematical fact that the inverse Wigner transform of a non-negative classical density need not yield a positive operator is, of course, well known and precisely the reason why $\mathcal{C}$ and $\mathcal{Q}$ do not coincide. The operational point, however, is that covariance based separability criteria, routinely deployed in continuous variable systems to certify entanglement, are blind to this non-positivity.

To visualize when classical mixtures appear entangled in this way, we analyze a tunable non-Gaussian example and track how the RS and PPT criteria respond. Consider the mixture of two displaced two-mode Gaussians
\begin{equation}
\label{eq:Gaussian_mixture_text}
    P(z) = \frac{1}{2} [G(z, \mu_+, \Sigma_0) + G(z, \mu_-, \Sigma_0)],
\end{equation}
where $z$ collects the phase-space coordinates, the two Gaussians $G$ share the same internal covariance $\Sigma_0$, and their means are oppositely displaced in position, i.e. $\mu_\pm = (\pm d, \mp d, 0, 0)$. For $d=0$, the distribution reduces to a single Gaussian, but for $d>0$ the mixture becomes non-Gaussian. The covariance matrix $\Sigma(d)$ of the mixture then combines the internal covariance $\Sigma_0$ with the additional spread introduced by mixing two displaced components. Although the explicit expression is given in Appendix \ref{app:covariance}, the important point is that $\Sigma(d)$ can be tuned by varying the displacement.

We can now apply the RS and PPT tests by computing the smallest symplectic eigenvalues associated with $\Sigma(d)$ and $\Sigma^\Gamma(d)$, respectively. 
For the underlying Gaussian at $d=0$, the RS bound is violated, confirming that the corresponding operator is purely classical. As we increase the displacement $d$, however, the covariance crosses into the RS-allowed region while violating the PPT condition, which would indicate entanglement. Crucially, though, the Weyl-transformed operator associated with $P(z)$ remains non-positive throughout this region, as shown numerically in Appendix \ref{app:negativity}. Therefore, the apparent ``entangled" covariance does not come from any quantum state, but from a classical distribution whose Hilbert-space operator fails positivity.

This is precisely what we call \emph{representational entanglement}: a regime in which classical correlations, once recast in Hilbert space and viewed under restricted access, mimic the signatures of genuine quantum entanglement, without constituting a genuine quantum resource \cite{collins_classical_2002, santo_which_2025}. This effect arises because the observer is still limited to second-moment information. Within this restricted view, entirely classical mixtures can reproduce the same covariance signatures that, in the quantum formalism, would be interpreted as entanglement.

\section{Hybrid and Genuine Entanglement}
\begin{figure}[ht]
    \centering
    \includegraphics[width=1\linewidth]{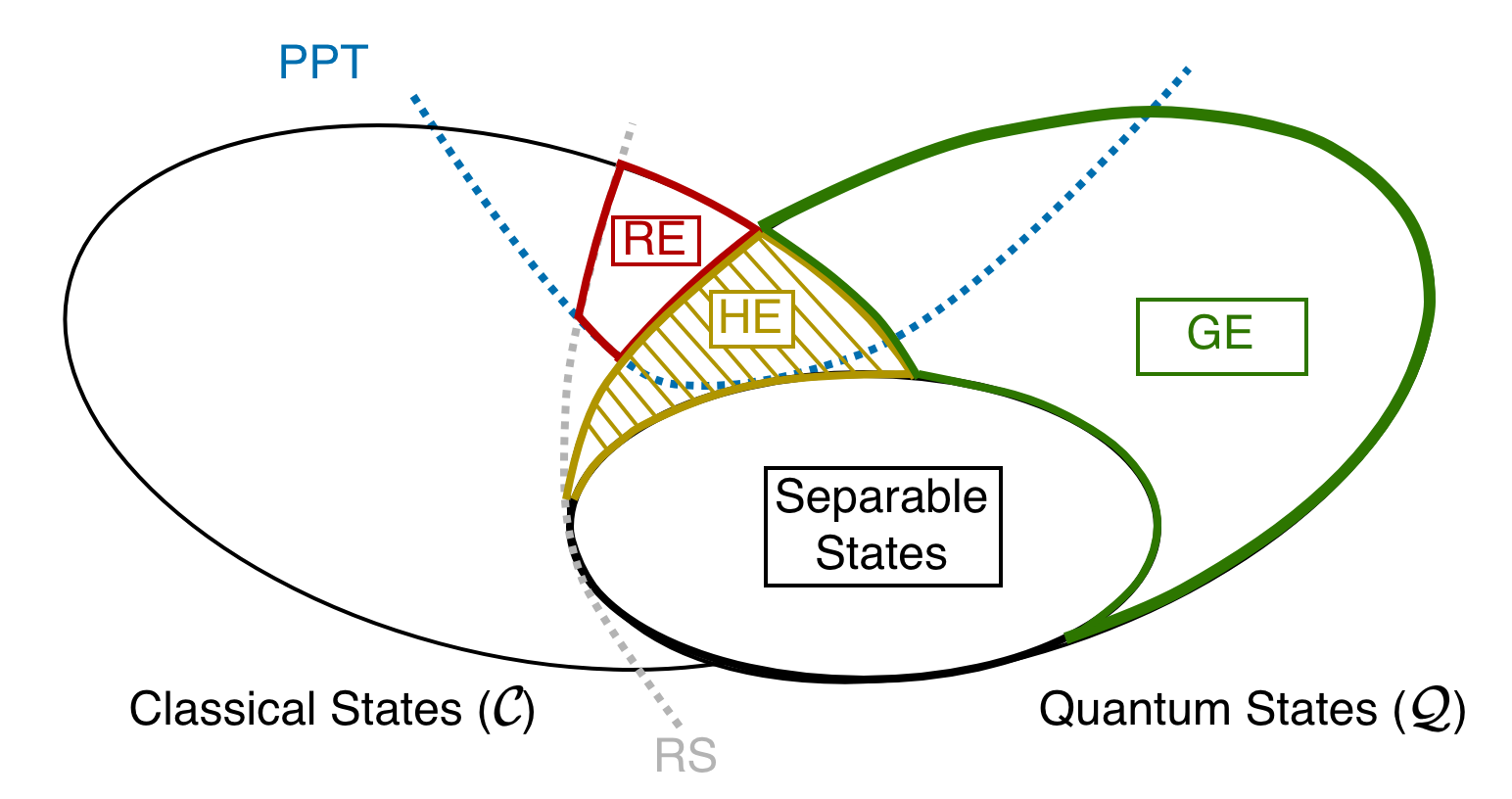}
    \caption{Three regimes of nonseparability: (i) RE: representational entanglement (non-positive), (ii) HE: hybrid entanglement (classically reproducible), and (iii) GE:  genuine entanglement (quantum-only). }
    \label{fig:entanglement}
\end{figure}

We now lift the observational restrictions of the previous section and allow the observer to test the operator-positivity condition (A), i.e., to determine whether the state corresponds to a positive semidefinite operator on Hilbert space. Operationally, we still assume that each experiment accesses only a single local phase-space (quadrature) setting per mode, as in standard continuous-variable experiments; however, by repeating the experiment for many settings, one can reconstruct the underlying operator via homodyne tomography. By sampling quadrature marginals over $\phi$
\begin{equation}
    x_\phi=\cos\phi\, q+\sin\phi\,\ p,\qquad \phi\in[0,\pi),
\end{equation}
one reconstructs $W(q,p)$ and the associated operator $\hat{\chi}$, which can then be tested for positivity (e.g. by diagonalization). In this way, one can arrive at the subset of states that satisfy (A) and are nonseparable according to the entanglement condition (B), while still being compatible with classical phase-space models in the sense that they admit a positive phase-space distribution. In other words, within the overlap \(C \cap Q\) the same positive Wigner function can be interpreted either as a classical probability density or as the Wigner function of a positive density operator, and for phase-space measurements, these two descriptions are operationally indistinguishable. We refer to nonseparability within this intersection as \emph{hybrid entanglement}, corresponding to the HE region in Fig.~\ref{fig:entanglement}. We shall further analyze the structure of such a set.

The Gaussian sector of this overlap has been characterised in detail in Ref.~\cite{bartlett_reconstruction_2012}, where it was shown that Gaussian quantum mechanics can be reconstructed from Liouville mechanics with an epistemic restriction on the available phase-space data; in particular, Gaussian entangled states populate $\mathcal{C}\cap\mathcal{Q}$, since their separability is fully characterised at the covariance level and they remain operationally classical under quadrature measurements. The relation between Wigner-function positivity and Gaussianity for single-mode mixed states has likewise been examined~\cite{Mandilara_2009, genoni_detecting_2013}. Our analysis differs from these prior results in two ways. First, the relevant setting here is bipartite and entangled, so the question is not whether a single-mode mixed state admits a positive Wigner function, but whether the resulting two-mode operator is simultaneously positive, entangled, and Wigner-positive. Second, we connect this question to operational diagnostics used in laboratory entanglement certification, delineating the boundaries between $\mathcal{C}\setminus\mathcal{Q}$, $\mathcal{C}\cap\mathcal{Q}$, and $\mathcal{Q}\setminus\mathcal{C}$ in operational rather than purely mathematical terms.

To further explore this hybrid region, we shall go beyond Gaussian states. For pure states we know that Gaussian quantum states saturate the entire classical--quantum overlap $\mathcal{C}\cap \mathcal{Q}$, as their corresponding Wigner function is positive \cite{hudson_when_1974}. For mixed states, however, this is no longer the case. A simple example is the convex mixture
\begin{equation}
\label{eq:single_mode_mixt}
    \rho(\eta) = \eta \ket{0}\bra{0} + (1-\eta) \ket{1} \bra{1},
\end{equation}
where the vacuum state corresponds to a positive Wigner function and the first excited Fock state exhibits Wigner function negativity. As $\eta$ decreases, the contribution of the non-classical component increases, eventually driving the Wigner function negative for $ \eta<1/2$. 

Thus the boundary between the hybrid and purely quantum region is at $\eta = 1/2$: for $\eta \in [1/2, 1]$ the state is Wigner-positive (hence classically compatible), and for $\eta \in [1/2, 1)$ it is moreover non-Gaussian, reducing to the (Gaussian) vacuum only at the point $\eta = 1$. For $\eta <1/2$ the state necessarily lies within the genuinely quantum region.

This single-mode example shows that, for mixed states, the classical-quantum overlap $\mathcal{C} \cap \mathcal{Q}$ extends strictly beyond the Gaussian subset $G_+$: for $\eta \in [1/2, 1]$ the state admits a non-negative Wigner function everywhere, and for 
$\eta \in [1/2, 1)$ it is moreover non-Gaussian. To obtain an entangled state in this overlap, we now embed $\rho(\eta)$ into a simple two-mode setting. This can be done by considering the following two-mode state $\rho_\text{in}$ = $\rho(\eta) \otimes \ket{0}\bra{0}$, subjected to a  balanced beamsplitter transformation and arriving at the following state
\begin{equation}
\begin{split}
    \rho_{AB}(\eta) &= \eta \ket{0,0} \bra{0,0} + (1-\eta) \ket{\psi_+} \bra{\psi_+}, \\ \text{with} \quad &\ket{\psi_+} = \frac{\ket{1,0} + \ket{0,1}}{\sqrt{2}}.
\end{split}
\end{equation}
At the phase-space level, a passive linear-optical transformation like a beamsplitter is just a symplectic rotation of the quadratures; thus, it cannot create or remove Wigner function negativity. Hence the two-mode Wigner function of $\rho_{AB}(\eta)$ is everywhere non-negative for the same parameter range $\eta \in [1/2, 1]$, in which the original single mode mixture $\rho(\eta)$ is Wigner-positive. In this range, $\rho_{AB}(\eta)$ is entangled for all $0 \leq \eta < 1$, and becomes separable only in the trivial limit $\eta \rightarrow 1$. Combining these facts, we obtain an illustrative example of \emph{hybrid entanglement}: for 
\begin{equation}
    \eta \in [1/2, 1) 
\end{equation}
the state $\rho_{AB}(\eta)$ is non-Gaussian and entangled, yet still admits a positive Wigner representation and hence admits a classical phase-space model. In the geometry of Fig. \ref{fig:entanglement}, it occupies the \emph{HE} region inside the overlap $\mathcal{C} \cap \mathcal{Q}$.

Another example is our displaced two-mode Gaussian mixture of Eq. \eqref{eq:Gaussian_mixture_text}, which also provides an example of a state in the hybrid regime. For suitable choices of intra-mode variances and inter-mode correlations (for instance, $s_q = s_p = 1$, $k_q = 0.3$, $k_p = -0.8$), the resulting mixture is non-Gaussian, satisfies RS and violates PPT at the covariance level, and its Weyl-transformed operator is numerically found to be non-negative up to numerical precision. Detailed calculations for the beamsplitter example can be found in Appendix \ref{app:hybrid_ent}.

These examples show that the hybrid region is nonempty and can be populated by both Gaussian and non-Gaussian states. The final step in our hierarchy is the domain of \emph{genuine entanglement (GE)}: here the reconstructed state is positive semidefinite yet exhibits Wigner-function negativity. States in this region (Fig.~\ref{fig:entanglement}) admit no classical phase-space description with an everywhere nonnegative distribution, and therefore cannot be reproduced by any classical model under the same measurement access.

\section{Conclusions and Outlook}
Our analysis shows that entanglement is not a unique feature of quantum theory, but can arise as a representational artifact or be mimicked by classical correlations in certain regimes. 
These facts give rise to an interesting discussion: if an experiment and the corresponding phase-space data analysis place the reconstructed state in the HE region, is the underlying system quantum or classical? From quadrature statistics alone, the answer is, in general, ambiguous, and additional criteria are required. One natural way to break this degeneracy is to probe the system under dynamics that go beyond quadratic Hamiltonians (i.e., genuinely non-Gaussian unitaries such as Kerr-type dynamics~\cite{oliva_quantum_2019}).

A second, more directly measurement-based route is to access regions of phase space where the underlying Wigner function might be negative. This is achieved by direct phase-space sampling via displaced photon-number parity, as proposed by Banaszek and W\'odkiewicz~\cite{Banaszek_1996, Banaszek_1999}. The measured quantity at phase-space point $(q,p)$ is, up to normalisation, $W(q,p)$ itself and any strictly negative value certifies that the state lies outside $\mathcal{C}$ and therefore outside the hybrid region, settling the ambiguity in favour of a genuinely quantum origin.

These distinctions come with a concrete hierarchy of experimental resources. Covariance data acquired from local quadrature measurements suffice to evaluate the RS and PPT criteria but cannot, by themselves, distinguish any of the three regimes. Full homodyne tomography gives access to the reconstructed operator and its spectrum, and thereby excludes (or reveals) representational entanglement, at a cost that grows with the Fock-space cutoff required for faithful reconstruction. Separating hybrid from genuine entanglement demands access to nonclassical signatures beyond quadratures: direct phase-space sampling via displaced photon-number parity~\cite{Banaszek_1996, Banaszek_1999}, or interrogation under non-Gaussian dynamics. In the optomechanical and gravitational settings discussed below, parity readout of massive mechanical modes is presently out of reach, so covariance-only certification is the norm, which is why the caution advocated here is of practical rather than merely conceptual relevance.

One may further ask whether hybrid-entangled states can arise dynamically from a physically plausible classical model. At the level of channels the answer is affirmative, as the displaced Gaussian mixture of Eq.~\eqref{eq:Gaussian_mixture_text} is prepared from a Gaussian state in $\mathcal{C} \cap \mathcal{Q}$ by a random displacement channel in which the two-mode displacement vector $\delta = (d,-d,0,0)^T$ is applied with sign $+$ or $-$ according to a shared classical random bit, i.e., by Gaussian unitaries conditioned on classical randomness, so that the entire preparation acts on the Wigner function as a classical stochastic map \cite{weedbrook_gaussian_2012}. More generally, any evolution whose generator acts on the Wigner function as a genuine Fokker--Planck operator (containing at most second order derivatives) maps non-negative distributions to non-negative distributions and therefore preserves $\mathcal{C}$ \cite{risken_fokker-planck_1996}. If such dynamics admits a hybrid-entangled steady state, the correlations mimicking entanglement are generated entirely by a classically simulable process. However, there is a structural restriction, as Lindblad generators whose Wigner representation is exactly of Fokker--Planck form are the quadratic ones, with linear drift and constant diffusion, and their steady states are Gaussian~\cite{gardiner2004quantum, serafini_quantum_2023}. Non-Gaussian hybrid-entangled states such as $\rho_{AB}(\eta)$ therefore cannot occur as exact steady states of such dynamics, although the truncated Wigner approximation~\cite{polkovnikov_phase_2010} may still give an approximate classical description in intermediate regimes. Conversely, leaving $\mathcal{C}$ requires Wigner flow containing higher order derivative terms, which is exactly the role played by the non-Gaussian unitaries described above.

Another interesting point was provided in Ref. \cite{santo_which_2025}, where it was emphasised that, although many quantum features—including entanglement—can already emerge in classical frameworks, it is ultimately the (in)compatibility between physical observables that fundamentally distinguishes classical from quantum physics. For example, to violate a Bell inequality, nonseparability alone is insufficient; one must measure incompatible observables. Here, we have shown that such incompatibility—specifically between position and momentum measurements—can be statistically emulated at the classical level simply by restricting an observer to a single phase-space observable per experimental run. Remarkably, this reverses the narrative: incompatibility should be viewed as a necessary but not sufficient condition for detecting genuine quantumness. Our examples show that incompatibility is a reliable signature of quantumness only when paired with a particular class of genuinely quantum states (exhibiting Wigner-function negativity) or with genuinely non-Gaussian dynamics that drive hybrid-entangled states out of the classical–quantum overlap into the GE regime.

A similar caution applies, more specifically, to the continuous-variable family of gravity-mediated entanglement (GME) proposals~\cite{Krisnanda_2020, datta2021signatures, kumar_continuous-variable_2023}, in which entanglement is inferred from second-order moments of the centre-of-mass quadratures of two masses and certified through covariance witnesses such as Duan--Simon or Robertson--Schr\"odinger. These schemes are directly susceptible to the representational and hybrid regimes identified in this work, as a violation of a Gaussian witness at the covariance level does not, on its own, certify either operator positivity or Wigner negativity of the underlying state. We therefore suggest that any covariance-based, continuous variable GME analysis be complemented by an explicit positivity check on the reconstructed operator and, where feasible, by direct phase-space sampling along the lines discussed above, so as to rule out the possibility that the observed correlations are reproducible by a classical phase-space mediator. Such an analysis is carried out in Ref.~\cite{schlegel2026gravitationalentanglementoptomechanicsdistinguishing},
where the standard optomechanical regime of Gaussian states, gravitational
coupling truncated at second order, and quadrature readout is shown to fall
entirely within the hybrid overlap, while the cubic term in the Newtonian
potential drives the quantum model into the genuinely entangled region and the
classical model into the representational one, at a quantified experimental
cost.

\emph{Acknowledgments.--}We warmly thank Nicolas Gisin, Tomasz Paterek, and Ankit Kumar for their insightful comments and helpful discussions on this work. This research was funded in whole, or in part, by the Austrian Science Fund (FWF) [10.55776/F71], [10.55776/P36994] and [10.55776/COE1] and the European Union – NextGenerationEU.  FDS acknowledges support from the Austrian Science Fund (FWF) through an Erwin Schr\"odinger Fellowship  [10.55776/J4699]. For open access purposes, the author(s) has applied a CC BY public copyright license to any author accepted manuscript version arising from this submission.

\appendix
\section{Gaussian Mixture} 
\subsection{Symplectic eigenvalues}\label{app:covariance}
The mixture of two Gaussians $G$ in Eq. \eqref{eq:Gaussian_mixture_text} yields a non-Gaussian state where the displacements cancel on average but contribute additional variance to the covariance matrix. To construct the covariance matrix $\Sigma$ of the mixture, we recall that the covariance of a mixed state consists of two contributions: the \emph{internal variance} within each component and the \emph{variance of the component means}. For our mixture centered around $\mu_\pm = ( \pm d, \mp d, 0, 0)$, both Gaussians share the same internal covariance $\Sigma_0$, while their means are displaced symmetrically around the origin. The overall covariance is therefore
\begin{equation}
    \Sigma = \Sigma_0 + \text{Cov}[\mu_\pm] = \Sigma_0 + \frac{1}{4}( \mu_+ - \mu_-)(\mu_+ - \mu_-)^T.
\end{equation}
Intuitively, the second term represents the additional spread originating by mixing two displaced components. Writing this out explicitly, for $z= (q_1, q_2, p_1, p_2)$, with intra-mode variances $s_q$, $s_p$ and inter-mode correlations $k_q$, $k_p$ we obtain 
\begin{equation}
\label{eq:cov_matr}
    \Sigma = \begin{pmatrix} s_q+ d^2 & k_q - d^2 & 0 & 0\\
    k_q - d^2 & s_q + d^2 & 0 & 0\\
    0 & 0 & s_p & k_p\\
    0 & 0 & k_p & s_p
    \end{pmatrix}.
\end{equation}
Note that for $d=0$ the mixture reduces to a single Gaussian state, where RS and PPT are both necessary and sufficient. Furthermore, any two mode covariance matrix can be brought into this form via local single-mode symplectic transformations. 

\begin{figure}[t]
    \centering
    \includegraphics[width=\linewidth]{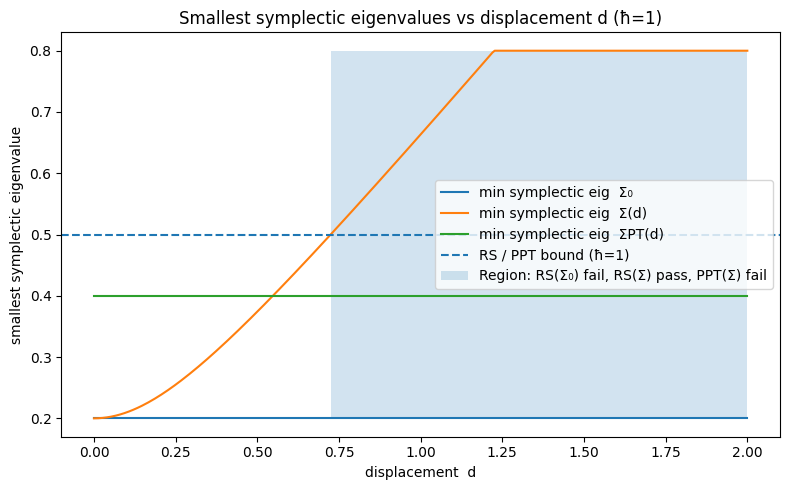}
    \caption{Smallest symplectic eigenvalues of the covariance matrix $\Sigma(d)$ 
    as a function of displacement $d$ (with $\hbar=1$). 
    The RS bound (dashed line) certifies physicality, while violation of PPT (green line below $1/2$) would normally indicate entanglement. 
    In the highlighted region, the covariance suggests a valid entangled state, but the underlying operator is non-positive, illustrating representational entanglement. Covariance-based analysis would misdiagnose entanglement, but the operator spectrum reveals non-positivity. With the parameter values $s_q = 0.5$, $s_p = 0.5$, $k_q = 0.3$ and $k_p = 0.3$.}
    \label{fig:symplectic}
\end{figure}

To calculate the symplectic eigenvalues of the different covariance matrices, it is easiest to find the eigenvalues of the matrix $i\Omega \Sigma$ and take their absolute values. For our quadrature-ordering the symplectic form is given by $\Omega =  \left(\begin{smallmatrix} 0 & \mathbb{I} \\ -\mathbb{I} & 0\end{smallmatrix} \right)$. Let us denote the smallest symplectic eigenvalue of $\Sigma$ as $\nu$ and the smallest symplectic eigenvalue of $\Sigma^\Gamma$ as $\tilde{\nu}$. This makes the RS condition of Eq. \eqref{eq:RS} and the PPT criterion from Eq. \eqref{eq:PPT}
\begin{align}
    \nu \geq \frac{\hbar}{2},\\
    \tilde{\nu} \geq \frac{\hbar}{2},
\end{align}
respectively. The symplectic eigenvalues of the total displaced Gaussian mixture of Eq. \eqref{eq:cov_matr} are given by
\begin{align}
    &\nu_1 = \sqrt{k_p k_q + k_p s_q + k_q s_p + s_p s_q},\\
   & \nu_2 =  \sqrt{-2d^2k_p + 2d^2s_p + k_p k_q - k_p s_q - k_q s_p + s_q s_p}.
\end{align}
Similarly, the eigenvalues of the partially transposed covariance can be calculated. The partial transposition on the covariance matrix simply leads to a flip of the sign on the momentum coordinate of the second system ($p_2 \rightarrow -p_2$), making the eigenvalues
\begin{align}
    &\tilde{\nu}_1 = \sqrt{-k_p k_q - k_p s_q + k_q s_p + s_p s_q},\\
    &\tilde{\nu}_2 =  \sqrt{2d^2k_p + 2d^2 s_p - k_p k_q + k_p s_q - k_q s_p + s_p s_q}.
\end{align}
The smallest of these eigenvalues, are then plotted against the displacement $d$ in Fig. \ref{fig:symplectic}, where the parameters take the values $s_q = 0.5$, $s_p = 0.5$, $k_q = 0.3$ and $k_p = 0.3$.

\subsection{Negativity of the Hilbert-space operator} \label{app:negativity}
\begin{figure}[t]
    \centering
    \includegraphics[width=1\linewidth]{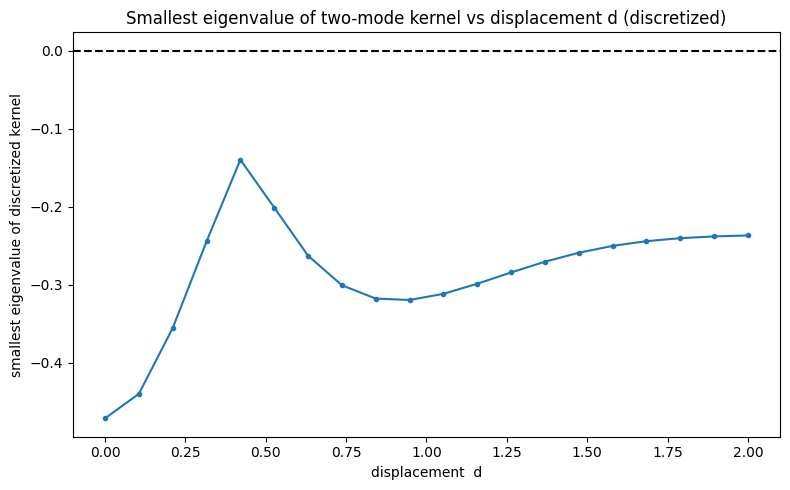}
    \caption{Smallest eigenvalue of the Weyl-transformed kernel $K(x,x')$ for the displaced Gaussian mixture $P(z)$ of Eq. \eqref{eq:Gaussian_mixture_text}, evaluated on a discretized position grid. For all displacements $d$, the kernel exhibits negative eigenvalues, confirming non-positivity of the corresponding Hilbert-space operator and thus the non-physical nature of the apparent entanglement.}
    \label{fig:negativity}
\end{figure}

To calculate negativity, we are required to take the Weyl transform $\hat{\rho}_P$ of our displaced Gaussian mixture of Eq. \eqref{eq:Gaussian_mixture_text}. Each Gaussian can be separately transformed via the following expression for its matrix elements
\begin{equation}
    \bra{x}\hat{\rho}_P\ket{x'} =\int d^2p P(m,p) e^{\frac{i}{\hbar} \Delta^T p},
\end{equation}
with $m = (x+x')/2$ and $\Delta = x-x'$. Since the covariance matrix $\Sigma_0 = \left(\begin{smallmatrix} Q_0 & 0 \\ 0 & P_0\end{smallmatrix} \right)$ of each Gaussian is block diagonal, we can simply pull the x-dependent part out of the integral
\begin{equation}
\begin{gathered}
    \bra{x}\hat{\rho}_P\ket{x'} = \frac{1}{(2\pi)^2 \sqrt{\det Q_0 \det P_0}}\\ \times \exp \left[  - \frac{1}{2} (m-\bar{q})^T Q_0^{-1} (m- \bar{q})\right] I(\Delta)
\end{gathered}
\end{equation}
and recognize the integral to be of standard Gaussian form 
\begin{equation}
    I(\Delta) = \int d^2p \, e^{-\frac{1}{2}p^T P_0^{-1}p + \frac{i}{\hbar}\Delta^T p},
\end{equation}
which can be computed to be 
\begin{equation}
    I(\Delta) = (2\pi) \sqrt{\det P_0} \exp \left[  - \frac{1}{2\hbar^2} \Delta^T P_0 \Delta\right].
\end{equation}
This gives us the final matrix element 
\begin{equation}
\begin{gathered}
        K(x,x', \mu_\pm, \Sigma_0) = \frac{1}{(2\pi)\sqrt{\det Q_0}}\\ \times \exp \left[ - \frac{1}{2} (m - \bar{q}_\pm)^T Q_0^{-1} (m- \bar{q}_\pm)  \right]  \\ \times \exp \left[ - \frac{1}{2\hbar^2}  \Delta^T P_0 \Delta\right],
\end{gathered}
\end{equation}
where $\bar{q}_\pm = (\pm d, \mp d)$. This position space kernel can then be discretized on a lattice. When performing this discretization on a finite grid, each matrix element $K_{ij} = K(X_i, X_j) (\Delta x)^2$ includes the configuration-space measure to approximate the continuum operator. The eigenvalue sign is unaffected by this scaling, but normalization and convergence improve. Choose, for instance, the parameter values $s_q = 0.5, s_p=0.5, k_q = 0.3, k_p = 0.3$ and numerically build the matrix on a grid ranging from $-8.0$ to $8.0$ with 50 lattice points along each axis. Finally, we can vary the displacement parameter from 0 to 2.0 and evaluate the smallest kernel eigenvalue at some of these discretized points. We deduce that the configuration space matrix remains negative for all the discussed displacement values, as seen in Fig. \ref{fig:negativity}.

\begin{figure}
    \centering
    \includegraphics[width=1\linewidth]{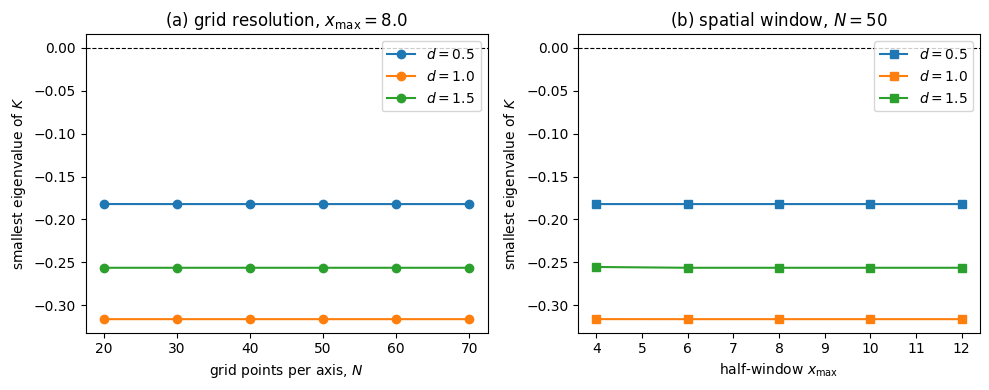}
    \caption{Convergence analysis for the displaced Gaussian mixture kernel for representative displacements $d\in\{0.5,1.0,1.5\}$. In the left panel we vary the grid resolution for a fixed spatial window, while the right panel varies the spatial window for a fixed number of discretization points.}
    \label{fig:convergence}
\end{figure}

To verify that the negative smallest eigenvalue of the discretised kernel $K(x,x')$ is not an artefact of the discretisation, we performed a convergence analysis in two 
independent directions: (i) refining the grid resolution $N$ at fixed spatial window $x_{\max}=8$, and (ii) enlarging the window $x_{\max}$ at fixed $N=50$. The results, shown in 
Fig.~\ref{fig:convergence} for three representative displacements $d\in\{0.5,1.0,1.5\}$, confirm that the smallest eigenvalue (a) approaches a stable negative value as $N$ increases beyond $
\sim 40$, and (b) is essentially independent of $x_{\max}$ for $x_{\max}\gtrsim 6$. The sign of the smallest eigenvalue is robust under both refinements, ruling out a numerical origin of the 
non-positivity reported in Fig.~\ref{fig:negativity}.

The same numerical procedure can be used to identify parameter choices of the displaced two-mode Gaussian mixture of Eq. \eqref{eq:Gaussian_mixture_text} that lie in the hybrid region. For instance, taking $s_q = s_p = 1$ and $k_q = 0.3$, $k_p = -0.8$ we find that the smallest eigenvalue of the Weyl-transformed kernel remains positive (any residual negativity is at the level of $10^{-16}$), while the state satisfies RS and violates covariance-based PPT. Since the underlying phase-space distribution is a classical Gaussian mixture, its Wigner function is manifestly non-negative. This confirms the existence of continuous-variable hybrid-entangled states consistent with the examples discussed above.

\section{Hybrid beamsplitter state}
\label{app:hybrid_ent}
We start from the single-mode mixed state of Eq. \eqref{eq:single_mode_mixt}, and prepare a two mode input state
\begin{equation}
    \rho_\text{in}(\eta) = \rho(\eta) \otimes \ket{0}\bra{0},
\end{equation}
where the second mode is in the vacuum. A balanced beamsplitter $U_{BS}$ acts on the relevant Fock states as
\begin{equation}
    U_{BS} \ket{0,0} = \ket{0,0}, \quad U_{BS} \ket{1,0} = \ket{\psi_+} := \frac{\ket{1,0} + \ket{0,1}}{\sqrt{2}}.
\end{equation}
The output state in the Fock basis $\{\ket{00}, \ket{01}, \ket{10}, \ket{11}\}$, is therefore
\begin{equation}
    \rho_{AB}(\eta) = \begin{pmatrix}
        \eta & 0 & 0 & 0\\
        0 & \frac{1}{2} - \frac{\eta}{2} & \frac{1}{2} - \frac{\eta}{2} & 0\\
        0 & \frac{1}{2} - \frac{\eta}{2} & \frac{1}{2} - \frac{\eta}{2} & 0\\
        0 & 0 & 0 & 0
    \end{pmatrix}.
\end{equation}
Partial transposition with respect to subsystem $B$ amounts to transposing the matrix elements that connect $\ket{0,1}$ and $\ket{1,0}$. This yields
\begin{equation}
    \rho_{AB}^{PT}(\eta) = \begin{pmatrix}
        \eta & 0 & 0 & \frac{1}{2} - \frac{\eta}{2}\\
        0 & \frac{1}{2} - \frac{\eta}{2} & 0 & 0\\
        0 & 0 & \frac{1}{2} - \frac{\eta}{2} & 0\\
        \frac{1}{2} - \frac{\eta}{2} & 0 & 0 & 0
    \end{pmatrix},
\end{equation}
where the matrix is block diagonal, with the eigenvalues
\begin{align}
    \lambda_{1,2}(\eta) &= \frac{1}{2}\left( \eta \pm \sqrt{2\eta^2-2\eta+1} \right),\\
    \lambda_{3,4}(\eta) &= \frac{1-\eta}{2} \geq 0 .
\end{align}
We find that 
\begin{equation}
    \lambda_1(\eta)= \frac{1}{2}\left( \eta - \sqrt{2\eta^2-2\eta+1} \right) < 0 \quad \text{for} \quad 0 \leq \eta < 1,
\end{equation}
meaning that the partial transpose is negative for all $\eta < 1$ and only becomes positive semidefinite in the trivial limit $\eta \rightarrow 1$. By the PPT criterion $\rho_{AB}(\eta)$ is entangled for any nonzero weight of the single-photon component. Combining this with the Wigner function positivity discussed above, one finds that for 
\begin{equation}
    \eta \in \left[1/2, 1\right),
\end{equation}
the state is simultaneously entangled and Wigner positive, and thus provides a concrete example of \emph{hybrid entanglement} within the $\mathcal{C} \cap \mathcal{Q}$ overlap.

Finally, let us discuss why the beamsplitter transformation preserves the positivity (or negativity) of the Wigner function. A linear-optical transformation, such as the beamsplitter, is generated by a quadratic Hamiltonian and hence corresponds to a Gaussian unitary. Therefore, at the level of canonical operators, it implements a symplectic orthogonal transformation $S$ on the quadrature vectors
\begin{equation}
    z_\text{out} = S \,z_\text{in},
\end{equation}
making the Wigner function after the beamsplitter
\begin{equation}
    W_\text{out}(z) = W_\text{in}(S^{-1}z),
\end{equation}
thus the beamsplitter does not change the value taken by the Wigner function, but only relabels the phase space coordinates. In our case the Wigner function factorizes as 
\begin{equation}
    W_\text{in} = W_{\rho(\eta)}(z_1) W_0(z_2),
\end{equation}
with $W_0 \geq 0$ everywhere for the vacuum. The sign of $W_\text{in}$ is therefore completely determined by $W_{\rho(\eta)}$. As discussed in the main text $W_{\rho(\eta)}$ is everywhere nonnegative for $\eta \geq 1/2$.


\begin{thebibliography}{48}%
\makeatletter
\providecommand \@ifxundefined [1]{%
 \@ifx{#1\undefined}
}%
\providecommand \@ifnum [1]{%
 \ifnum #1\expandafter \@firstoftwo
 \else \expandafter \@secondoftwo
 \fi
}%
\providecommand \@ifx [1]{%
 \ifx #1\expandafter \@firstoftwo
 \else \expandafter \@secondoftwo
 \fi
}%
\providecommand \natexlab [1]{#1}%
\providecommand \enquote  [1]{``#1''}%
\providecommand \bibnamefont  [1]{#1}%
\providecommand \bibfnamefont [1]{#1}%
\providecommand \citenamefont [1]{#1}%
\providecommand \href@noop [0]{\@secondoftwo}%
\providecommand \href [0]{\begingroup \@sanitize@url \@href}%
\providecommand \@href[1]{\@@startlink{#1}\@@href}%
\providecommand \@@href[1]{\endgroup#1\@@endlink}%
\providecommand \@sanitize@url [0]{\catcode `\\12\catcode `\$12\catcode `\&12\catcode `\#12\catcode `\^12\catcode `\_12\catcode `\%12\relax}%
\providecommand \@@startlink[1]{}%
\providecommand \@@endlink[0]{}%
\providecommand \url  [0]{\begingroup\@sanitize@url \@url }%
\providecommand \@url [1]{\endgroup\@href {#1}{\urlprefix }}%
\providecommand \urlprefix  [0]{URL }%
\providecommand \Eprint [0]{\href }%
\providecommand \doibase [0]{https://doi.org/}%
\providecommand \selectlanguage [0]{\@gobble}%
\providecommand \bibinfo  [0]{\@secondoftwo}%
\providecommand \bibfield  [0]{\@secondoftwo}%
\providecommand \translation [1]{[#1]}%
\providecommand \BibitemOpen [0]{}%
\providecommand \bibitemStop [0]{}%
\providecommand \bibitemNoStop [0]{.\EOS\space}%
\providecommand \EOS [0]{\spacefactor3000\relax}%
\providecommand \BibitemShut  [1]{\csname bibitem#1\endcsname}%
\let\auto@bib@innerbib\@empty
\bibitem [{\citenamefont {Schr{\"o}dinger}(1935)}]{schrodinger1935discussion}%
  \BibitemOpen
  \bibfield  {author} {\bibinfo {author} {\bibfnamefont {E.}~\bibnamefont {Schr{\"o}dinger}},\ }in\ \href@noop {} {\emph {\bibinfo {booktitle} {Mathematical Proceedings of the Cambridge Philosophical Society}}},\ Vol.~\bibinfo {volume} {31}\ (\bibinfo {organization} {Cambridge University Press},\ \bibinfo {year} {1935})\ pp.\ \bibinfo {pages} {555--563}\BibitemShut {NoStop}%
\bibitem [{\citenamefont {Spekkens}(2007)}]{spekkens_defense_2007}%
  \BibitemOpen
  \bibfield  {author} {\bibinfo {author} {\bibfnamefont {R.~W.}\ \bibnamefont {Spekkens}},\ }\bibfield  {title} {\emph {\bibinfo {title} {In defense of the epistemic view of quantum states: a toy theory}},\ }\href {https://doi.org/10.1103/PhysRevA.75.032110} {\bibfield  {journal} {\bibinfo  {journal} {Physical Review A}\ }\textbf {\bibinfo {volume} {75}},\ \bibinfo {pages} {032110} (\bibinfo {year} {2007})},\ \bibinfo {note} {arXiv:quant-ph/0401052}\BibitemShut {NoStop}%
\bibitem [{\citenamefont {Catani}\ \emph {et~al.}(2022)\citenamefont {Catani}, \citenamefont {Leifer}, \citenamefont {Scala}, \citenamefont {Schmid},\ and\ \citenamefont {Spekkens}}]{catani1}%
  \BibitemOpen
  \bibfield  {author} {\bibinfo {author} {\bibfnamefont {L.}~\bibnamefont {Catani}}, \bibinfo {author} {\bibfnamefont {M.}~\bibnamefont {Leifer}}, \bibinfo {author} {\bibfnamefont {G.}~\bibnamefont {Scala}}, \bibinfo {author} {\bibfnamefont {D.}~\bibnamefont {Schmid}},\ and\ \bibinfo {author} {\bibfnamefont {R.~W.}\ \bibnamefont {Spekkens}},\ }\bibfield  {title} {\emph {\bibinfo {title} {What is nonclassical about uncertainty relations?}},\ }\href {https://doi.org/10.1103/PhysRevLett.129.240401} {\bibfield  {journal} {\bibinfo  {journal} {Phys. Rev. Lett.}\ }\textbf {\bibinfo {volume} {129}},\ \bibinfo {pages} {240401} (\bibinfo {year} {2022})}\BibitemShut {NoStop}%
\bibitem [{\citenamefont {Catani}\ \emph {et~al.}(2023{\natexlab{a}})\citenamefont {Catani}, \citenamefont {Leifer}, \citenamefont {Schmid},\ and\ \citenamefont {Spekkens}}]{catani2023interference}%
  \BibitemOpen
  \bibfield  {author} {\bibinfo {author} {\bibfnamefont {L.}~\bibnamefont {Catani}}, \bibinfo {author} {\bibfnamefont {M.}~\bibnamefont {Leifer}}, \bibinfo {author} {\bibfnamefont {D.}~\bibnamefont {Schmid}},\ and\ \bibinfo {author} {\bibfnamefont {R.~W.}\ \bibnamefont {Spekkens}},\ }\bibfield  {title} {\emph {\bibinfo {title} {Why interference phenomena do not capture the essence of quantum theory}},\ }\href {https://doi.org/https://doi.org/10.22331/q-2023-09-25-1119} {\bibfield  {journal} {\bibinfo  {journal} {Quantum}\ }\textbf {\bibinfo {volume} {7}},\ \bibinfo {pages} {1119} (\bibinfo {year} {2023}{\natexlab{a}})}\BibitemShut {NoStop}%
\bibitem [{\citenamefont {Catani}\ \emph {et~al.}(2023{\natexlab{b}})\citenamefont {Catani}, \citenamefont {Leifer}, \citenamefont {Scala}, \citenamefont {Schmid},\ and\ \citenamefont {Spekkens}}]{catani2023aspects}%
  \BibitemOpen
  \bibfield  {author} {\bibinfo {author} {\bibfnamefont {L.}~\bibnamefont {Catani}}, \bibinfo {author} {\bibfnamefont {M.}~\bibnamefont {Leifer}}, \bibinfo {author} {\bibfnamefont {G.}~\bibnamefont {Scala}}, \bibinfo {author} {\bibfnamefont {D.}~\bibnamefont {Schmid}},\ and\ \bibinfo {author} {\bibfnamefont {R.~W.}\ \bibnamefont {Spekkens}},\ }\bibfield  {title} {\emph {\bibinfo {title} {Aspects of the phenomenology of interference that are genuinely nonclassical}},\ }\href {https://doi.org/https://doi.org/10.1103/PhysRevA.108.022207} {\bibfield  {journal} {\bibinfo  {journal} {Physical Review A}\ }\textbf {\bibinfo {volume} {108}},\ \bibinfo {pages} {022207} (\bibinfo {year} {2023}{\natexlab{b}})}\BibitemShut {NoStop}%
\bibitem [{\citenamefont {Fankhauser}\ \emph {et~al.}(2025)\citenamefont {Fankhauser}, \citenamefont {Gonda},\ and\ \citenamefont {De~les Coves}}]{Fankhauser_2025}%
  \BibitemOpen
  \bibfield  {author} {\bibinfo {author} {\bibfnamefont {J.}~\bibnamefont {Fankhauser}}, \bibinfo {author} {\bibfnamefont {T.}~\bibnamefont {Gonda}},\ and\ \bibinfo {author} {\bibfnamefont {G.}~\bibnamefont {De~les Coves}},\ }\bibfield  {title} {\emph {\bibinfo {title} {Epistemic horizons from deterministic laws: lessons from a nomic toy theory}},\ }\bibfield  {journal} {\bibinfo  {journal} {Synthese}\ }\textbf {\bibinfo {volume} {205}},\ \href {https://doi.org/10.1007/s11229-024-04852-0} {10.1007/s11229-024-04852-0} (\bibinfo {year} {2025})\BibitemShut {NoStop}%
\bibitem [{\citenamefont {Del~Santo}\ and\ \citenamefont {Gisin}(2019)}]{del2019physics}%
  \BibitemOpen
  \bibfield  {author} {\bibinfo {author} {\bibfnamefont {F.}~\bibnamefont {Del~Santo}}\ and\ \bibinfo {author} {\bibfnamefont {N.}~\bibnamefont {Gisin}},\ }\bibfield  {title} {\emph {\bibinfo {title} {Physics without determinism: Alternative interpretations of classical physics}},\ }\href@noop {} {\bibfield  {journal} {\bibinfo  {journal} {Physical Review A}\ }\textbf {\bibinfo {volume} {100}},\ \bibinfo {pages} {062107} (\bibinfo {year} {2019})}\BibitemShut {NoStop}%
\bibitem [{\citenamefont {Santo}\ and\ \citenamefont {Gisin}(2025)}]{santo_which_2025}%
  \BibitemOpen
  \bibfield  {author} {\bibinfo {author} {\bibfnamefont {F.~D.}\ \bibnamefont {Santo}}\ and\ \bibinfo {author} {\bibfnamefont {N.}~\bibnamefont {Gisin}},\ }\bibfield  {title} {\emph {\bibinfo {title} {Which features of quantum physics are not fundamentally quantum but are due to indeterminism?}},\ }\href {https://doi.org/10.22331/q-2025-04-03-1686} {\bibfield  {journal} {\bibinfo  {journal} {Quantum}\ }\textbf {\bibinfo {volume} {9}},\ \bibinfo {pages} {1686} (\bibinfo {year} {2025})},\ \bibinfo {note} {arXiv:2409.10601 [quant-ph]}\BibitemShut {NoStop}%
\bibitem [{\citenamefont {Chiribella}\ \emph {et~al.}(2024)\citenamefont {Chiribella}, \citenamefont {Giannelli},\ and\ \citenamefont {Scandolo}}]{chiribella_bell_2024}%
  \BibitemOpen
  \bibfield  {author} {\bibinfo {author} {\bibfnamefont {G.}~\bibnamefont {Chiribella}}, \bibinfo {author} {\bibfnamefont {L.}~\bibnamefont {Giannelli}},\ and\ \bibinfo {author} {\bibfnamefont {C.~M.}\ \bibnamefont {Scandolo}},\ }\bibfield  {title} {\emph {\bibinfo {title} {Bell {Nonlocality} in {Classical} {Systems} {Coexisting} with {Other} {System} {Types}}},\ }\href {https://doi.org/10.1103/PhysRevLett.132.190201} {\bibfield  {journal} {\bibinfo  {journal} {Physical Review Letters}\ }\textbf {\bibinfo {volume} {132}},\ \bibinfo {pages} {190201} (\bibinfo {year} {2024})}\BibitemShut {NoStop}%
\bibitem [{\citenamefont {D'Ariano}\ \emph {et~al.}(2020)\citenamefont {D'Ariano}, \citenamefont {Erba},\ and\ \citenamefont {Perinotti}}]{dariano_classical_2020}%
  \BibitemOpen
  \bibfield  {author} {\bibinfo {author} {\bibfnamefont {G.~M.}\ \bibnamefont {D'Ariano}}, \bibinfo {author} {\bibfnamefont {M.}~\bibnamefont {Erba}},\ and\ \bibinfo {author} {\bibfnamefont {P.}~\bibnamefont {Perinotti}},\ }\bibfield  {title} {\emph {\bibinfo {title} {Classical theories with entanglement}},\ }\href {https://doi.org/10.1103/PhysRevA.101.042118} {\bibfield  {journal} {\bibinfo  {journal} {Physical Review A}\ }\textbf {\bibinfo {volume} {101}},\ \bibinfo {pages} {042118} (\bibinfo {year} {2020})}\BibitemShut {NoStop}%
\bibitem [{\citenamefont {Spreeuw}(1998)}]{spreeuw_classical_1998}%
  \BibitemOpen
  \bibfield  {author} {\bibinfo {author} {\bibfnamefont {R.~J.~C.}\ \bibnamefont {Spreeuw}},\ }\bibfield  {title} {\emph {\bibinfo {title} {A {Classical} {Analogy} of {Entanglement}}},\ }\href {https://doi.org/10.1023/A:1018703709245} {\bibfield  {journal} {\bibinfo  {journal} {Foundations of Physics}\ }\textbf {\bibinfo {volume} {28}},\ \bibinfo {pages} {361--374} (\bibinfo {year} {1998})}\BibitemShut {NoStop}%
\bibitem [{\citenamefont {Aiello}\ \emph {et~al.}(2015)\citenamefont {Aiello}, \citenamefont {T{\"o}ppel}, \citenamefont {Marquardt}, \citenamefont {Giacobino},\ and\ \citenamefont {Leuchs}}]{aiello_quantum_2015}%
  \BibitemOpen
  \bibfield  {author} {\bibinfo {author} {\bibfnamefont {A.}~\bibnamefont {Aiello}}, \bibinfo {author} {\bibfnamefont {F.}~\bibnamefont {T{\"o}ppel}}, \bibinfo {author} {\bibfnamefont {C.}~\bibnamefont {Marquardt}}, \bibinfo {author} {\bibfnamefont {E.}~\bibnamefont {Giacobino}},\ and\ \bibinfo {author} {\bibfnamefont {G.}~\bibnamefont {Leuchs}},\ }\bibfield  {title} {\emph {\bibinfo {title} {Quantum like nonseparable structures in optical beams}},\ }\href {https://doi.org/10.1088/1367-2630/17/4/043024} {\bibfield  {journal} {\bibinfo  {journal} {New Journal of Physics}\ }\textbf {\bibinfo {volume} {17}},\ \bibinfo {pages} {043024} (\bibinfo {year} {2015})}\BibitemShut {NoStop}%
\bibitem [{\citenamefont {Forbes}\ \emph {et~al.}(2019)\citenamefont {Forbes}, \citenamefont {Aiello},\ and\ \citenamefont {Ndagano}}]{forbes_chapter_2019}%
  \BibitemOpen
  \bibfield  {author} {\bibinfo {author} {\bibfnamefont {A.}~\bibnamefont {Forbes}}, \bibinfo {author} {\bibfnamefont {A.}~\bibnamefont {Aiello}},\ and\ \bibinfo {author} {\bibfnamefont {B.}~\bibnamefont {Ndagano}},\ }in\ \href {https://doi.org/10.1016/bs.po.2018.11.001} {\emph {\bibinfo {booktitle} {Progress in {Optics}}}},\ Vol.~\bibinfo {volume} {64},\ \bibinfo {editor} {edited by\ \bibinfo {editor} {\bibfnamefont {T.~D.}\ \bibnamefont {Visser}}}\ (\bibinfo  {publisher} {Elsevier},\ \bibinfo {year} {2019})\ pp.\ \bibinfo {pages} {99--153}\BibitemShut {NoStop}%
\bibitem [{\citenamefont {Karimi}\ and\ \citenamefont {Boyd}(2015)}]{karimi_classical_2015}%
  \BibitemOpen
  \bibfield  {author} {\bibinfo {author} {\bibfnamefont {E.}~\bibnamefont {Karimi}}\ and\ \bibinfo {author} {\bibfnamefont {R.~W.}\ \bibnamefont {Boyd}},\ }\bibfield  {title} {\emph {\bibinfo {title} {Classical entanglement?}},\ }\href {https://doi.org/10.1126/science.aad7174} {\bibfield  {journal} {\bibinfo  {journal} {Science}\ }\textbf {\bibinfo {volume} {350}},\ \bibinfo {pages} {1172--1173} (\bibinfo {year} {2015})}\BibitemShut {NoStop}%
\bibitem [{\citenamefont {Korolkova}\ \emph {et~al.}(2024)\citenamefont {Korolkova}, \citenamefont {S{\'a}nchez-Soto},\ and\ \citenamefont {Leuchs}}]{korolkova_operational_2024}%
  \BibitemOpen
  \bibfield  {author} {\bibinfo {author} {\bibfnamefont {N.}~\bibnamefont {Korolkova}}, \bibinfo {author} {\bibfnamefont {L.}~\bibnamefont {S{\'a}nchez-Soto}},\ and\ \bibinfo {author} {\bibfnamefont {G.}~\bibnamefont {Leuchs}},\ }\bibfield  {title} {\emph {\bibinfo {title} {An operational distinction between quantum entanglement and classical non-separability}},\ }\href {https://doi.org/10.1098/rsta.2023.0342} {\bibfield  {journal} {\bibinfo  {journal} {Philosophical Transactions of the Royal Society A: Mathematical, Physical and Engineering Sciences}\ }\textbf {\bibinfo {volume} {382}},\ \bibinfo {pages} {20230342} (\bibinfo {year} {2024})}\BibitemShut {NoStop}%
\bibitem [{\citenamefont {Peres}(1997)}]{peres1997quantum}%
  \BibitemOpen
  \bibfield  {author} {\bibinfo {author} {\bibfnamefont {A.}~\bibnamefont {Peres}},\ }\href@noop {} {\emph {\bibinfo {title} {Quantum theory: concepts and methods}}},\ Vol.~\bibinfo {volume} {57}\ (\bibinfo  {publisher} {Springer},\ \bibinfo {year} {1997})\BibitemShut {NoStop}%
\bibitem [{\citenamefont {Gerry}\ and\ \citenamefont {Knight}(2023)}]{gerry_introductory_2023}%
  \BibitemOpen
  \bibfield  {author} {\bibinfo {author} {\bibfnamefont {C.~C.}\ \bibnamefont {Gerry}}\ and\ \bibinfo {author} {\bibfnamefont {P.~L.}\ \bibnamefont {Knight}},\ }\href@noop {} {\emph {\bibinfo {title} {Introductory {Quantum} {Optics}}}}\ (\bibinfo  {publisher} {Cambridge University Press},\ \bibinfo {year} {2023})\BibitemShut {NoStop}%
\bibitem [{\citenamefont {Duan}\ \emph {et~al.}(2000)\citenamefont {Duan}, \citenamefont {Giedke}, \citenamefont {Cirac},\ and\ \citenamefont {Zoller}}]{duan_inseparability_2000}%
  \BibitemOpen
  \bibfield  {author} {\bibinfo {author} {\bibfnamefont {L.-M.}\ \bibnamefont {Duan}}, \bibinfo {author} {\bibfnamefont {G.}~\bibnamefont {Giedke}}, \bibinfo {author} {\bibfnamefont {J.~I.}\ \bibnamefont {Cirac}},\ and\ \bibinfo {author} {\bibfnamefont {P.}~\bibnamefont {Zoller}},\ }\bibfield  {title} {\emph {\bibinfo {title} {Inseparability {Criterion} for {Continuous} {Variable} {Systems}}},\ }\href {https://doi.org/10.1103/PhysRevLett.84.2722} {\bibfield  {journal} {\bibinfo  {journal} {Physical Review Letters}\ }\textbf {\bibinfo {volume} {84}},\ \bibinfo {pages} {2722--2725} (\bibinfo {year} {2000})}\BibitemShut {NoStop}%
\bibitem [{\citenamefont {Serafini}(2023)}]{serafini_quantum_2023}%
  \BibitemOpen
  \bibfield  {author} {\bibinfo {author} {\bibfnamefont {A.}~\bibnamefont {Serafini}},\ }\href {https://doi.org/10.1201/9781003250975} {\emph {\bibinfo {title} {Quantum {Continuous} {Variables}: {A} {Primer} of {Theoretical} {Methods}}}},\ \bibinfo {edition} {2nd}\ ed.\ (\bibinfo  {publisher} {CRC Press},\ \bibinfo {address} {Boca Raton},\ \bibinfo {year} {2023})\BibitemShut {NoStop}%
\bibitem [{\citenamefont {Weedbrook}\ \emph {et~al.}(2012)\citenamefont {Weedbrook}, \citenamefont {Pirandola}, \citenamefont {Garcia-Patron}, \citenamefont {Cerf}, \citenamefont {Ralph}, \citenamefont {Shapiro},\ and\ \citenamefont {Lloyd}}]{weedbrook_gaussian_2012}%
  \BibitemOpen
  \bibfield  {author} {\bibinfo {author} {\bibfnamefont {C.}~\bibnamefont {Weedbrook}}, \bibinfo {author} {\bibfnamefont {S.}~\bibnamefont {Pirandola}}, \bibinfo {author} {\bibfnamefont {R.}~\bibnamefont {Garcia-Patron}}, \bibinfo {author} {\bibfnamefont {N.~J.}\ \bibnamefont {Cerf}}, \bibinfo {author} {\bibfnamefont {T.~C.}\ \bibnamefont {Ralph}}, \bibinfo {author} {\bibfnamefont {J.~H.}\ \bibnamefont {Shapiro}},\ and\ \bibinfo {author} {\bibfnamefont {S.}~\bibnamefont {Lloyd}},\ }\bibfield  {title} {\emph {\bibinfo {title} {Gaussian {Quantum} {Information}}},\ }\href {https://doi.org/10.1103/RevModPhys.84.621} {\bibfield  {journal} {\bibinfo  {journal} {Reviews of Modern Physics}\ }\textbf {\bibinfo {volume} {84}},\ \bibinfo {pages} {621--669} (\bibinfo {year} {2012})},\ \bibinfo {note} {arXiv:1110.3234 [quant-ph]}\BibitemShut {NoStop}%
\bibitem [{\citenamefont {Mandilara}\ \emph {et~al.}(2009)\citenamefont {Mandilara}, \citenamefont {Karpov},\ and\ \citenamefont {Cerf}}]{Mandilara_2009}%
  \BibitemOpen
  \bibfield  {author} {\bibinfo {author} {\bibfnamefont {A.}~\bibnamefont {Mandilara}}, \bibinfo {author} {\bibfnamefont {E.}~\bibnamefont {Karpov}},\ and\ \bibinfo {author} {\bibfnamefont {N.~J.}\ \bibnamefont {Cerf}},\ }\bibfield  {title} {\emph {\bibinfo {title} {Extending hudson's theorem to mixed quantum states}},\ }\bibfield  {journal} {\bibinfo  {journal} {Physical Review A}\ }\textbf {\bibinfo {volume} {79}},\ \href {https://doi.org/10.1103/physreva.79.062302} {10.1103/physreva.79.062302} (\bibinfo {year} {2009})\BibitemShut {NoStop}%
\bibitem [{\citenamefont {Genoni}\ \emph {et~al.}(2013)\citenamefont {Genoni}, \citenamefont {Palma}, \citenamefont {Tufarelli}, \citenamefont {Olivares}, \citenamefont {Kim},\ and\ \citenamefont {Paris}}]{genoni_detecting_2013}%
  \BibitemOpen
  \bibfield  {author} {\bibinfo {author} {\bibfnamefont {M.~G.}\ \bibnamefont {Genoni}}, \bibinfo {author} {\bibfnamefont {M.~L.}\ \bibnamefont {Palma}}, \bibinfo {author} {\bibfnamefont {T.}~\bibnamefont {Tufarelli}}, \bibinfo {author} {\bibfnamefont {S.}~\bibnamefont {Olivares}}, \bibinfo {author} {\bibfnamefont {M.~S.}\ \bibnamefont {Kim}},\ and\ \bibinfo {author} {\bibfnamefont {M.~G.~A.}\ \bibnamefont {Paris}},\ }\bibfield  {title} {\emph {\bibinfo {title} {Detecting quantum non-{Gaussianity} via the {Wigner} function}},\ }\href {https://doi.org/10.1103/PhysRevA.87.062104} {\bibfield  {journal} {\bibinfo  {journal} {Physical Review A}\ }\textbf {\bibinfo {volume} {87}},\ \bibinfo {pages} {062104} (\bibinfo {year} {2013})},\ \bibinfo {note} {arXiv:1304.3340 [quant-ph]}\BibitemShut {NoStop}%
\bibitem [{\citenamefont {Bartlett}\ \emph {et~al.}(2012)\citenamefont {Bartlett}, \citenamefont {Rudolph},\ and\ \citenamefont {Spekkens}}]{bartlett_reconstruction_2012}%
  \BibitemOpen
  \bibfield  {author} {\bibinfo {author} {\bibfnamefont {S.~D.}\ \bibnamefont {Bartlett}}, \bibinfo {author} {\bibfnamefont {T.}~\bibnamefont {Rudolph}},\ and\ \bibinfo {author} {\bibfnamefont {R.~W.}\ \bibnamefont {Spekkens}},\ }\bibfield  {title} {\emph {\bibinfo {title} {Reconstruction of {Gaussian} quantum mechanics from {Liouville} mechanics with an epistemic restriction}},\ }\href {https://doi.org/10.1103/PhysRevA.86.012103} {\bibfield  {journal} {\bibinfo  {journal} {Physical Review A}\ }\textbf {\bibinfo {volume} {86}},\ \bibinfo {pages} {012103} (\bibinfo {year} {2012})},\ \bibinfo {note} {arXiv:1111.5057 [quant-ph]}\BibitemShut {NoStop}%
\bibitem [{\citenamefont {Gomes}\ \emph {et~al.}(2009)\citenamefont {Gomes}, \citenamefont {Salles}, \citenamefont {Toscano}, \citenamefont {Souto~Ribeiro},\ and\ \citenamefont {Walborn}}]{gomes_quantum_2009}%
  \BibitemOpen
  \bibfield  {author} {\bibinfo {author} {\bibfnamefont {R.~M.}\ \bibnamefont {Gomes}}, \bibinfo {author} {\bibfnamefont {A.}~\bibnamefont {Salles}}, \bibinfo {author} {\bibfnamefont {F.}~\bibnamefont {Toscano}}, \bibinfo {author} {\bibfnamefont {P.~H.}\ \bibnamefont {Souto~Ribeiro}},\ and\ \bibinfo {author} {\bibfnamefont {S.~P.}\ \bibnamefont {Walborn}},\ }\bibfield  {title} {\emph {\bibinfo {title} {Quantum entanglement beyond {Gaussian} criteria}},\ }\href {https://doi.org/10.1073/pnas.0908329106} {\bibfield  {journal} {\bibinfo  {journal} {Proceedings of the National Academy of Sciences}\ }\textbf {\bibinfo {volume} {106}},\ \bibinfo {pages} {21517--21520} (\bibinfo {year} {2009})}\BibitemShut {NoStop}%
\bibitem [{\citenamefont {Hertz}\ \emph {et~al.}(2016)\citenamefont {Hertz}, \citenamefont {Karpov}, \citenamefont {Mandilara},\ and\ \citenamefont {Cerf}}]{hertz_detection_2016}%
  \BibitemOpen
  \bibfield  {author} {\bibinfo {author} {\bibfnamefont {A.}~\bibnamefont {Hertz}}, \bibinfo {author} {\bibfnamefont {E.}~\bibnamefont {Karpov}}, \bibinfo {author} {\bibfnamefont {A.}~\bibnamefont {Mandilara}},\ and\ \bibinfo {author} {\bibfnamefont {N.~J.}\ \bibnamefont {Cerf}},\ }\bibfield  {title} {\emph {\bibinfo {title} {Detection of non-{Gaussian} entangled states with an improved continuous-variable separability criterion}},\ }\href {https://doi.org/10.1103/PhysRevA.93.032330} {\bibfield  {journal} {\bibinfo  {journal} {Physical Review A}\ }\textbf {\bibinfo {volume} {93}},\ \bibinfo {pages} {032330} (\bibinfo {year} {2016})},\ \bibinfo {note} {arXiv:1511.06621 [quant-ph]}\BibitemShut {NoStop}%
\bibitem [{\citenamefont {Weyl}(1927)}]{weyl_quantenmechanik_1927}%
  \BibitemOpen
  \bibfield  {author} {\bibinfo {author} {\bibfnamefont {H.}~\bibnamefont {Weyl}},\ }\bibfield  {title} {\emph {\bibinfo {title} {Quantenmechanik und {Gruppentheorie}}},\ }\href {https://doi.org/10.1007/BF02055756} {\bibfield  {journal} {\bibinfo  {journal} {Zeitschrift f{\"u}r Physik}\ }\textbf {\bibinfo {volume} {46}},\ \bibinfo {pages} {1--46} (\bibinfo {year} {1927})}\BibitemShut {NoStop}%
\bibitem [{\citenamefont {Moyal}(1949)}]{moyal_quantum_1949}%
  \BibitemOpen
  \bibfield  {author} {\bibinfo {author} {\bibfnamefont {J.~E.}\ \bibnamefont {Moyal}},\ }\bibfield  {title} {\emph {\bibinfo {title} {Quantum mechanics as a statistical theory}},\ }\href {https://doi.org/10.1017/S0305004100000487} {\bibfield  {journal} {\bibinfo  {journal} {Mathematical Proceedings of the Cambridge Philosophical Society}\ }\textbf {\bibinfo {volume} {45}},\ \bibinfo {pages} {99--124} (\bibinfo {year} {1949})}\BibitemShut {NoStop}%
\bibitem [{\citenamefont {Case}(2008)}]{case_wigner_2008}%
  \BibitemOpen
  \bibfield  {author} {\bibinfo {author} {\bibfnamefont {W.~B.}\ \bibnamefont {Case}},\ }\bibfield  {title} {\emph {\bibinfo {title} {Wigner functions and {Weyl} transforms for pedestrians}},\ }\href {https://doi.org/10.1119/1.2957889} {\bibfield  {journal} {\bibinfo  {journal} {American Journal of Physics}\ }\textbf {\bibinfo {volume} {76}},\ \bibinfo {pages} {937--946} (\bibinfo {year} {2008})}\BibitemShut {NoStop}%
\bibitem [{\citenamefont {Hillery}\ \emph {et~al.}(1984)\citenamefont {Hillery}, \citenamefont {O'Connell}, \citenamefont {Scully},\ and\ \citenamefont {Wigner}}]{hillery_distribution_1984}%
  \BibitemOpen
  \bibfield  {author} {\bibinfo {author} {\bibfnamefont {M.}~\bibnamefont {Hillery}}, \bibinfo {author} {\bibfnamefont {R.~F.}\ \bibnamefont {O'Connell}}, \bibinfo {author} {\bibfnamefont {M.~O.}\ \bibnamefont {Scully}},\ and\ \bibinfo {author} {\bibfnamefont {E.~P.}\ \bibnamefont {Wigner}},\ }\bibfield  {title} {\emph {\bibinfo {title} {Distribution functions in physics: {Fundamentals}}},\ }\href {https://doi.org/10.1016/0370-1573(84)90160-1} {\bibfield  {journal} {\bibinfo  {journal} {Physics Reports}\ }\textbf {\bibinfo {volume} {106}},\ \bibinfo {pages} {121--167} (\bibinfo {year} {1984})}\BibitemShut {NoStop}%
\bibitem [{\citenamefont {Kenfack}\ and\ \citenamefont {Zyczkowski}(2004)}]{kenfack_negativity_2004}%
  \BibitemOpen
  \bibfield  {author} {\bibinfo {author} {\bibfnamefont {A.}~\bibnamefont {Kenfack}}\ and\ \bibinfo {author} {\bibfnamefont {K.}~\bibnamefont {Zyczkowski}},\ }\bibfield  {title} {\emph {\bibinfo {title} {Negativity of the {Wigner} function as an indicator of nonclassicality}},\ }\href {https://doi.org/10.1088/1464-4266/6/10/003} {\bibfield  {journal} {\bibinfo  {journal} {Journal of Optics B: Quantum and Semiclassical Optics}\ }\textbf {\bibinfo {volume} {6}},\ \bibinfo {pages} {396--404} (\bibinfo {year} {2004})},\ \bibinfo {note} {arXiv:quant-ph/0406015}\BibitemShut {NoStop}%
\bibitem [{Note1()}]{Note1}%
  \BibitemOpen
  \bibinfo {note} {$\protect \hat {\Delta }(q,p) = \protect \int _{-\infty }^{\infty } ds\protect \tmspace +\thickmuskip {.2777em} e^{\protect \frac {i}{\hbar }ps}\protect \, \mathinner {|{q+\protect \tfrac {s}{2}}\rangle }\mathinner {\langle {q-\protect \tfrac {s}{2}}|}$}\BibitemShut {NoStop}%
\bibitem [{\citenamefont {Stratonovich}(1957)}]{stratonovich1957distributions}%
  \BibitemOpen
  \bibfield  {author} {\bibinfo {author} {\bibfnamefont {R.~L.}\ \bibnamefont {Stratonovich}},\ }\bibfield  {title} {\emph {\bibinfo {title} {On distributions in representation space}},\ }\href@noop {} {\bibfield  {journal} {\bibinfo  {journal} {Soviet Physics JETP-USSR}\ }\textbf {\bibinfo {volume} {4}},\ \bibinfo {pages} {891--898} (\bibinfo {year} {1957})}\BibitemShut {NoStop}%
\bibitem [{\citenamefont {Ockeloen-Korppi}\ \emph {et~al.}(2018)\citenamefont {Ockeloen-Korppi}, \citenamefont {Damsk{\"a}gg}, \citenamefont {Pirkkalainen}, \citenamefont {Asjad}, \citenamefont {Clerk}, \citenamefont {Massel}, \citenamefont {Woolley},\ and\ \citenamefont {Sillanp{\"a}{\"a}}}]{ockeloen-korppi_stabilized_2018}%
  \BibitemOpen
  \bibfield  {author} {\bibinfo {author} {\bibfnamefont {C.~F.}\ \bibnamefont {Ockeloen-Korppi}}, \bibinfo {author} {\bibfnamefont {E.}~\bibnamefont {Damsk{\"a}gg}}, \bibinfo {author} {\bibfnamefont {J.-M.}\ \bibnamefont {Pirkkalainen}}, \bibinfo {author} {\bibfnamefont {M.}~\bibnamefont {Asjad}}, \bibinfo {author} {\bibfnamefont {A.~A.}\ \bibnamefont {Clerk}}, \bibinfo {author} {\bibfnamefont {F.}~\bibnamefont {Massel}}, \bibinfo {author} {\bibfnamefont {M.~J.}\ \bibnamefont {Woolley}},\ and\ \bibinfo {author} {\bibfnamefont {M.~A.}\ \bibnamefont {Sillanp{\"a}{\"a}}},\ }\bibfield  {title} {\emph {\bibinfo {title} {Stabilized entanglement of massive mechanical oscillators}},\ }\href {https://doi.org/10.1038/s41586-018-0038-x} {\bibfield  {journal} {\bibinfo  {journal} {Nature}\ }\textbf {\bibinfo {volume} {556}},\ \bibinfo {pages} {478--482} (\bibinfo {year} {2018})}\BibitemShut {NoStop}%
\bibitem [{\citenamefont {Aspelmeyer}\ \emph {et~al.}(2014)\citenamefont {Aspelmeyer}, \citenamefont {Kippenberg},\ and\ \citenamefont {Marquardt}}]{aspelmeyer_cavity_2014}%
  \BibitemOpen
  \bibfield  {author} {\bibinfo {author} {\bibfnamefont {M.}~\bibnamefont {Aspelmeyer}}, \bibinfo {author} {\bibfnamefont {T.~J.}\ \bibnamefont {Kippenberg}},\ and\ \bibinfo {author} {\bibfnamefont {F.}~\bibnamefont {Marquardt}},\ }\bibfield  {title} {\emph {\bibinfo {title} {Cavity optomechanics}},\ }\href {https://doi.org/10.1103/RevModPhys.86.1391} {\bibfield  {journal} {\bibinfo  {journal} {Reviews of Modern Physics}\ }\textbf {\bibinfo {volume} {86}},\ \bibinfo {pages} {1391--1452} (\bibinfo {year} {2014})}\BibitemShut {NoStop}%
\bibitem [{\citenamefont {Simon}(2000)}]{simon_peres-horodecki_2000}%
  \BibitemOpen
  \bibfield  {author} {\bibinfo {author} {\bibfnamefont {R.}~\bibnamefont {Simon}},\ }\bibfield  {title} {\emph {\bibinfo {title} {Peres-{Horodecki} separability criterion for continuous variable systems}},\ }\href {https://doi.org/10.1103/PhysRevLett.84.2726} {\bibfield  {journal} {\bibinfo  {journal} {Physical Review Letters}\ }\textbf {\bibinfo {volume} {84}},\ \bibinfo {pages} {2726--2729} (\bibinfo {year} {2000})},\ \bibinfo {note} {arXiv:quant-ph/9909044}\BibitemShut {NoStop}%
\bibitem [{\citenamefont {Adesso}\ and\ \citenamefont {Illuminati}(2007)}]{adesso_entanglement_2007}%
  \BibitemOpen
  \bibfield  {author} {\bibinfo {author} {\bibfnamefont {G.}~\bibnamefont {Adesso}}\ and\ \bibinfo {author} {\bibfnamefont {F.}~\bibnamefont {Illuminati}},\ }\bibfield  {title} {\emph {\bibinfo {title} {Entanglement in continuous variable systems: {Recent} advances and current perspectives}},\ }\href {https://doi.org/10.1088/1751-8113/40/28/S01} {\bibfield  {journal} {\bibinfo  {journal} {Journal of Physics A: Mathematical and Theoretical}\ }\textbf {\bibinfo {volume} {40}},\ \bibinfo {pages} {7821--7880} (\bibinfo {year} {2007})},\ \bibinfo {note} {arXiv:quant-ph/0701221}\BibitemShut {NoStop}%
\bibitem [{\citenamefont {Collins}\ and\ \citenamefont {Popescu}(2002)}]{collins_classical_2002}%
  \BibitemOpen
  \bibfield  {author} {\bibinfo {author} {\bibfnamefont {D.}~\bibnamefont {Collins}}\ and\ \bibinfo {author} {\bibfnamefont {S.}~\bibnamefont {Popescu}},\ }\bibfield  {title} {\emph {\bibinfo {title} {Classical analog of entanglement}},\ }\href {https://doi.org/10.1103/PhysRevA.65.032321} {\bibfield  {journal} {\bibinfo  {journal} {Physical Review A}\ }\textbf {\bibinfo {volume} {65}},\ \bibinfo {pages} {032321} (\bibinfo {year} {2002})}\BibitemShut {NoStop}%
\bibitem [{\citenamefont {Hudson}(1974)}]{hudson_when_1974}%
  \BibitemOpen
  \bibfield  {author} {\bibinfo {author} {\bibfnamefont {R.~L.}\ \bibnamefont {Hudson}},\ }\bibfield  {title} {\emph {\bibinfo {title} {When is the wigner quasi-probability density non-negative?}},\ }\href {https://doi.org/10.1016/0034-4877(74)90007-X} {\bibfield  {journal} {\bibinfo  {journal} {Reports on Mathematical Physics}\ }\textbf {\bibinfo {volume} {6}},\ \bibinfo {pages} {249--252} (\bibinfo {year} {1974})}\BibitemShut {NoStop}%
\bibitem [{\citenamefont {Oliva}\ and\ \citenamefont {Steuernagel}(2019)}]{oliva_quantum_2019}%
  \BibitemOpen
  \bibfield  {author} {\bibinfo {author} {\bibfnamefont {M.}~\bibnamefont {Oliva}}\ and\ \bibinfo {author} {\bibfnamefont {O.}~\bibnamefont {Steuernagel}},\ }\bibfield  {title} {\emph {\bibinfo {title} {Quantum {Kerr} oscillators' evolution in phase space: {Wigner} current, symmetries, shear suppression, and special states}},\ }\href {https://doi.org/10.1103/PhysRevA.99.032104} {\bibfield  {journal} {\bibinfo  {journal} {Physical Review A}\ }\textbf {\bibinfo {volume} {99}},\ \bibinfo {pages} {032104} (\bibinfo {year} {2019})}\BibitemShut {NoStop}%
\bibitem [{\citenamefont {Banaszek}\ and\ \citenamefont {W{\'o}dkiewicz}(1996)}]{Banaszek_1996}%
  \BibitemOpen
  \bibfield  {author} {\bibinfo {author} {\bibfnamefont {K.}~\bibnamefont {Banaszek}}\ and\ \bibinfo {author} {\bibfnamefont {K.}~\bibnamefont {W{\'o}dkiewicz}},\ }\bibfield  {title} {\emph {\bibinfo {title} {Direct probing of quantum phase space by photon counting}},\ }\href {https://doi.org/10.1103/physrevlett.76.4344} {\bibfield  {journal} {\bibinfo  {journal} {Physical Review Letters}\ }\textbf {\bibinfo {volume} {76}},\ \bibinfo {pages} {4344--4347} (\bibinfo {year} {1996})}\BibitemShut {NoStop}%
\bibitem [{\citenamefont {Banaszek}\ \emph {et~al.}(1999)\citenamefont {Banaszek}, \citenamefont {Radzewicz}, \citenamefont {W{\'o}dkiewicz},\ and\ \citenamefont {Krasi{\'n}ski}}]{Banaszek_1999}%
  \BibitemOpen
  \bibfield  {author} {\bibinfo {author} {\bibfnamefont {K.}~\bibnamefont {Banaszek}}, \bibinfo {author} {\bibfnamefont {C.}~\bibnamefont {Radzewicz}}, \bibinfo {author} {\bibfnamefont {K.}~\bibnamefont {W{\'o}dkiewicz}},\ and\ \bibinfo {author} {\bibfnamefont {J.~S.}\ \bibnamefont {Krasi{\'n}ski}},\ }\bibfield  {title} {\emph {\bibinfo {title} {Direct measurement of the wigner function by photon counting}},\ }\href {https://doi.org/10.1103/physreva.60.674} {\bibfield  {journal} {\bibinfo  {journal} {Physical Review A}\ }\textbf {\bibinfo {volume} {60}},\ \bibinfo {pages} {674--677} (\bibinfo {year} {1999})}\BibitemShut {NoStop}%
\bibitem [{\citenamefont {Risken}(1996)}]{risken_fokker-planck_1996}%
  \BibitemOpen
  \bibfield  {author} {\bibinfo {author} {\bibfnamefont {H.}~\bibnamefont {Risken}},\ }in\ \href {https://doi.org/10.1007/978-3-642-61544-3_4} {\emph {\bibinfo {booktitle} {The {Fokker}-{Planck} {Equation}: {Methods} of {Solution} and {Applications}}}},\ \bibinfo {editor} {edited by\ \bibinfo {editor} {\bibfnamefont {H.}~\bibnamefont {Risken}}}\ (\bibinfo  {publisher} {Springer},\ \bibinfo {address} {Berlin, Heidelberg},\ \bibinfo {year} {1996})\ pp.\ \bibinfo {pages} {63--95}\BibitemShut {NoStop}%
\bibitem [{\citenamefont {Gardiner}\ and\ \citenamefont {Zoller}(2004)}]{gardiner2004quantum}%
  \BibitemOpen
  \bibfield  {author} {\bibinfo {author} {\bibfnamefont {C.}~\bibnamefont {Gardiner}}\ and\ \bibinfo {author} {\bibfnamefont {P.}~\bibnamefont {Zoller}},\ }\href@noop {} {\emph {\bibinfo {title} {Quantum noise: a handbook of Markovian and non-Markovian quantum stochastic methods with applications to quantum optics}}}\ (\bibinfo  {publisher} {Springer Science \& Business Media},\ \bibinfo {year} {2004})\BibitemShut {NoStop}%
\bibitem [{\citenamefont {Polkovnikov}(2010)}]{polkovnikov_phase_2010}%
  \BibitemOpen
  \bibfield  {author} {\bibinfo {author} {\bibfnamefont {A.}~\bibnamefont {Polkovnikov}},\ }\bibfield  {title} {\emph {\bibinfo {title} {Phase space representation of quantum dynamics}},\ }\href {https://doi.org/10.1016/j.aop.2010.02.006} {\bibfield  {journal} {\bibinfo  {journal} {Annals of Physics}\ }\textbf {\bibinfo {volume} {325}},\ \bibinfo {pages} {1790--1852} (\bibinfo {year} {2010})},\ \bibinfo {note} {arXiv:0905.3384 [cond-mat]}\BibitemShut {NoStop}%
\bibitem [{\citenamefont {Krisnanda}\ \emph {et~al.}(2020)\citenamefont {Krisnanda}, \citenamefont {Tham}, \citenamefont {Paternostro},\ and\ \citenamefont {Paterek}}]{Krisnanda_2020}%
  \BibitemOpen
  \bibfield  {author} {\bibinfo {author} {\bibfnamefont {T.}~\bibnamefont {Krisnanda}}, \bibinfo {author} {\bibfnamefont {G.~Y.}\ \bibnamefont {Tham}}, \bibinfo {author} {\bibfnamefont {M.}~\bibnamefont {Paternostro}},\ and\ \bibinfo {author} {\bibfnamefont {T.}~\bibnamefont {Paterek}},\ }\bibfield  {title} {\emph {\bibinfo {title} {Observable quantum entanglement due to gravity}},\ }\bibfield  {journal} {\bibinfo  {journal} {npj Quantum Information}\ }\textbf {\bibinfo {volume} {6}},\ \href {https://doi.org/10.1038/s41534-020-0243-y} {10.1038/s41534-020-0243-y} (\bibinfo {year} {2020})\BibitemShut {NoStop}%
\bibitem [{\citenamefont {Datta}\ and\ \citenamefont {Miao}(2021)}]{datta2021signatures}%
  \BibitemOpen
  \bibfield  {author} {\bibinfo {author} {\bibfnamefont {A.}~\bibnamefont {Datta}}\ and\ \bibinfo {author} {\bibfnamefont {H.}~\bibnamefont {Miao}},\ }\bibfield  {title} {\emph {\bibinfo {title} {Signatures of the quantum nature of gravity in the differential motion of two masses}},\ }\href@noop {} {\bibfield  {journal} {\bibinfo  {journal} {Quantum Science \& Technology}\ }\textbf {\bibinfo {volume} {6}},\ \bibinfo {pages} {045014} (\bibinfo {year} {2021})}\BibitemShut {NoStop}%
\bibitem [{\citenamefont {Kumar}\ \emph {et~al.}(2023)\citenamefont {Kumar}, \citenamefont {Krisnanda}, \citenamefont {Arumugam},\ and\ \citenamefont {Paterek}}]{kumar_continuous-variable_2023}%
  \BibitemOpen
  \bibfield  {author} {\bibinfo {author} {\bibfnamefont {A.}~\bibnamefont {Kumar}}, \bibinfo {author} {\bibfnamefont {T.}~\bibnamefont {Krisnanda}}, \bibinfo {author} {\bibfnamefont {P.}~\bibnamefont {Arumugam}},\ and\ \bibinfo {author} {\bibfnamefont {T.}~\bibnamefont {Paterek}},\ }\bibfield  {title} {\emph {\bibinfo {title} {Continuous-{Variable} {Entanglement} through {Central} {Forces}: {Application} to {Gravity} between {Quantum} {Masses}}},\ }\href {https://doi.org/10.22331/q-2023-05-15-1008} {\bibfield  {journal} {\bibinfo  {journal} {Quantum}\ }\textbf {\bibinfo {volume} {7}},\ \bibinfo {pages} {1008} (\bibinfo {year} {2023})}\BibitemShut {NoStop}%
\bibitem [{\citenamefont {Schlegel}\ \emph {et~al.}(2026)\citenamefont {Schlegel}, \citenamefont {Kumar}, \citenamefont {Paterek},\ and\ \citenamefont {Daki{\'c}}}]{schlegel2026gravitationalentanglementoptomechanicsdistinguishing}%
  \BibitemOpen
  \bibfield  {author} {\bibinfo {author} {\bibfnamefont {S.}~\bibnamefont {Schlegel}}, \bibinfo {author} {\bibfnamefont {A.}~\bibnamefont {Kumar}}, \bibinfo {author} {\bibfnamefont {T.}~\bibnamefont {Paterek}},\ and\ \bibinfo {author} {\bibfnamefont {B.}~\bibnamefont {Daki{\'c}}},\ }\href {https://arxiv.org/abs/2605.20330} {\bibinfo {title} {Gravitational entanglement in optomechanics: Distinguishing classical and quantum models}} (\bibinfo {year} {2026}),\ \Eprint {https://arxiv.org/abs/2605.20330} {arXiv:2605.20330 [quant-ph]} \BibitemShut {NoStop}%
\end{thebibliography}
\end{document}